\documentclass[preprint,12pt]{elsarticle}

\usepackage{amssymb}
\usepackage{amsmath}

\usepackage{tikz}
\usepackage{subcaption}
\usepackage{url}

\newcommand{\p}{\rho}

\newcommand{\R}{\mathbb{R}}                
\newcommand{\rhoSet}{\mathcal{P}}
\newcommand{\QSet}{\mathcal{Q}}

\newtheorem{proposition}{Proposition}
\newtheorem{theorem}{Theorem}
\newtheorem{corollary}{Corollary}

\newcommand{\Normal}[2]{\mathcal{N} \left(#1, #2 \right)}
\newcommand{\vectwod}[2]{\begin{pmatrix}
#1 \\
#2
\end{pmatrix}}
\newcommand{\E}[1]{E\left[ #1 \right]}
\newcommand{\var}[1]{\mathrm{Var}(#1)}
\newcommand{\covar}[2]{\mathrm{Cov}(#1, #2)}
\newcommand{\partialDer}[2]{\frac{\partial #1}{\partial #2}}
\newcommand{\derivative}[2]{\frac{d #1}{d #2}}
\newcommand{\sgn}[1]{\text{sgn}\left( #1 \right)}
\newcommand{\simiid}{\overset{iid}{\sim}}

\journal{IJAR}

\begin{document}

\begin{frontmatter}


\title{Sensitivity Analysis to Unobserved Confounding with Copula-based Normalizing Flows}

\author[a]{Sourabh Balgi} 
\author[b]{Marc Braun} 
\author[c]{Jose M. Pe\~na} 
\author[d]{Adel Daoud} 

\affiliation[a]{organization={IDA, Linköping University},
            addressline={}, 
            city={Linköping},
            postcode={58183}, 
            state={},
            country={Sweden}}

\affiliation[b]{organization={IDA, Linköping University},
            addressline={}, 
            city={Linköping},
            postcode={58183}, 
            state={},
            country={Sweden}}

\affiliation[c]{organization={IDA, Linköping University},
            addressline={}, 
            city={Linköping},
            postcode={58183}, 
            state={},
            country={Sweden}}

\affiliation[d]{organization={IEI, Linköping University},
            addressline={}, 
            city={Linköping},
            postcode={58183}, 
            state={},
            country={Sweden}}

\begin{abstract}

We propose a novel method for sensitivity analysis to unobserved confounding in causal inference. The method builds on a copula-based causal graphical normalizing flow that we term $\rho$-GNF, where $\rho \in [-1,+1]$ is the sensitivity parameter. The parameter represents the non-causal association between exposure and outcome due to unobserved confounding, which is modeled as a Gaussian copula. In other words, the $\rho$-GNF enables scholars to estimate the average causal effect (ACE) as a function of $\rho$, accounting for various confounding strengths. The output of the $\rho$-GNF is what we term the $\rho_{curve}$, which provides the bounds for the ACE given an interval of assumed $\rho$ values. The $\rho_{curve}$ also enables scholars to identify the confounding strength required to nullify the ACE. We also propose a Bayesian version of our sensitivity analysis method. Assuming a prior over the sensitivity parameter $\rho$ enables us to derive the posterior distribution over the ACE, which enables us to derive credible intervals. Finally, leveraging on experiments from simulated and real-world data, we show the benefits of our sensitivity analysis method.

\end{abstract}


\end{frontmatter}


\section{Introduction}\label{sec:intro}

Epidemiologists, sociologists, economists, and other applied scientists, often leverage randomized controlled trials (RCTs), as these provide the safest methodological route to perform causal inference \citep{wright1921correlationcausation, fisher1936designofexperiments, cox1958planning, Imbens2015CIinSocialscience}. By randomizing which experimental subjects (e.g., people, villages, schools) should take the treatment and which subjects should abstain, an RCT ensures unconfoundedness (also known as ignorability or exchangeability): There is no unobserved common causes of the treatment and the outcome in the causal system under study. When unconfoundedness is satisfied, scholars can calculate the causal effect of interest from collected data; that means that the causal quantity is identifiable \citep{pearl2009causality, hernan2009ipw}.
However, despite the importance of RCTs, they remain infeasible for a slew of applied settings. They may be costly to implement (e.g., testing a population-wide medicine); they may be unethical (e.g., testing a new drug); or they may be impractical to implement (e.g., testing a social policy across the world). Therefore, applied researchers often rely on observational data -- which are often secondary data sources with no treatment randomization and where the experimenter had no control over the data generating process. Yet when using observational data, scholars make themselves susceptible for failing to satisfy the unconfoundedness assumption, even when some confounders are observed.

Because the unconfoundedness assumption is so critical and at the same time untestable in observational studies \citep{pearl2009causality, hernan2009ipw}, methodologists (statisticians, computer scientist, and others) have developed various frameworks for stress testing how causal effect estimates change under varying degree of unconfoundedness failure. These sorts of tests are named sensitivity analysis \citep{schlesselman1978assessing, manski1990sensitivitybounds, imbens2003sensitivity, brumback2004sensitivity, vanderweele2011bias}, also known as bias analysis in epidemiology \citep{cornfield1959smoking, cochran1973controlling, rothman2008modern, lash2009applying}.
Nonetheless, existing sensitivity analysis methods are limited in at least three ways, as we elaborate on below and in the next section. In this paper, we propose a new method to improve on these limitations, thereby moving the state-of-the-art forward. 

First, we propose a deep learning method for sensitivity analysis based on causal graphical normalizing flows \citep{wehenkel2020GNF,balgi2022cgnf,JavaloySV23}, because of their attractive properties of non-linearity and invertibility for counterfactual inference. Specifically, we combine these normalizing flows with the Gaussian copula \citep{nelsen2007introduction}, the most studied and popular elliptical copula, to model unobserved confounding. Hence, we aptly name the new model $\rho$-GNF, where $\rho \in [-1,+1]$ is the sensitivity parameter of the Gaussian copula that controls the degree of unconfoundedness between treatment and outcome. Unlike most sensitivity analysis methods where the sensitivity parameters are unbounded and thus difficult to specify and interpret, $\rho$ is bounded and has a clear interpretation. 

Second, we show that our $\rho$-GNF enables us to estimate the average causal effect (ACE) as a function of $\rho$. This function, which we call $\rho_{curve}$, enables us to identify the ACE bounds given a specific interval of $\rho$ that the domain expert considers appropriate. It also enables us to determine the confounding strength required to explain away the causal effect, which we call $\rho_{value}$. This is similar to the widely used E-value \citep{vanderweele2017sensitivity}. Unlike most sensitivity analysis methods, our method can be cast in a Bayesian setting by allowing the user to specify a prior distribution over $\rho$, to return the posterior distribution over the ACE and corresponding credible intervals.

Third, $\rho$-GNF accommodates both discrete and continuous outcomes. This is in contrast to most existing sensitivity analysis methods, which apply to only one type of outcome. For instance, the works \citep{robins1989analysis, manski1990sensitivitybounds, sjolander2020note, sjolander2021novel, pena2022simple} require the outcome variable being binary, while the works \citep{cinelli2019sensitivity, cinelli2020making} require the outcome to be continuous.

The remainder of the paper is structured as follows. After reviewing the related literature in Section \ref{sec:background}, we introduce $\rho$-GNF for sensitivity analysis in Section \ref{sec:rho-GNF}. Moreover, we introduce a Bayesian extension of it in Section \ref{sec:Bayesian}, and discuss suitable choices for prior distributions of $\rho$. In Section \ref{sec:experiments}, we present our results with simulated and real-world data under different settings of the outcome variable (i.e., continuous, binary and categorical), and compare them with the popular assumption free bounds \citep{robins1989analysis, manski1990sensitivitybounds}. Finally, in Section \ref{sec:conclusion}, we conclude by discussing the key contributions of our $\rho$-GNF method.

It is worth mentioning that this work is an extension of the conference contribution \citep{balgi2022rho}. Specifically, the extension consists of the Bayesian framework presented in Section \ref{sec:Bayesian}, its evaluation on simulated and real-world datasets in Sections \ref{sec:experiments simcont} and \ref{sec:real}, and the proofs of the formal results included in \ref{proposition} and \ref{theorem}.

\section{Related Works}\label{sec:background}

The literature on sensitivity analysis can be roughly categorized into two streams: (i) identify the bounds of the causal effect as functions of some sensitivity parameters that encode the strength of the unobserved confounders \citep{robins1989analysis, manski1990sensitivitybounds, vanderweele2011bias, ding2016sensitivity, sjolander2020note, sjolander2021novel, pena2022simple}, and (ii) identify how large the influence of the unobserved confounders must be to explain away the causal effect \citep{imbens2003sensitivity, vanderweele2017sensitivity, veitch2020sense, sjolander2022values}.
For instance, the works \citep{robins1989analysis, manski1990sensitivitybounds} provide assumption free bounds of the causal effect for binary outcomes, while more recent works extend the bounds to categorical outcomes \citep{ilse2021efficient}. Other methods provide bounds as functions of sensitivity parameters to be tuned by a domain expert \citep{sjolander2020note, sjolander2021novel, pena2022simple}. Others are limited to the linear setting \citep{cinelli2019sensitivity, cinelli2020making}. In contrast to the bounds stream, there are methods that fall under the explain-away stream. For example, the works \citep{imbens2003sensitivity, vanderweele2017sensitivity} reason on the minimum strength of the unmeasured confounder that is needed, conditional on the measured confounders, to explain away the causal effect. Similarly, the recently developed Austen plots identify the influence of the confounding needed to explain a specific amount of bias in the causal effect estimate \citep{veitch2020sense}. 

As shown above, there is a wide spectrum of sensitivity analysis methods. All certainly have unique advantages, but they also have limitations. For example, the methods in the works \citep{robins1989analysis, manski1990sensitivitybounds, sjolander2020note, sjolander2021novel, pena2022simple} require the outcome variable being binary. The methods in the work \citep{cinelli2019sensitivity, cinelli2020making} assume a linear model. The method in the work \citep{vanderweele2017sensitivity} may result in wider bounds than the assumption free bounds \citep{ioannidis2019limitations, sjolander2020note}, which are thus logically impossible. Some methods are exclusively suited for specific causal estimands, e.g., ACE or conditional ACE or mediation effects \citep{tchetgen2012semiparametric, lindmark2018sensitivity}. Some methods offer no sensitivity parameters to determine the confounding strength required to explain away the causal effect \citep{robins1989analysis, manski1990sensitivitybounds}. Others like those in the works \citep{vanderweele2017sensitivity, veitch2020sense} offer multiple parameters that are unbounded and hard to specify for the domain analyst \citep{ioannidis2019limitations}.

To summarize, even though there exist several sensitivity analysis frameworks, there is still a lack of unifying method that is flexible enough to suit many different types of observational data, with easy-to-use sensitivity parameters. Moreover, it is imperative to establish a method enabling researchers to specify distributional assumptions regarding the unobserved causes within the causal system under study. Such assumptions enable tighter ACE bounds, enhancing the certainty of the findings. Our $\rho$-GNF method targets all these lacks. Our method uses deep neural networks, allowing for maximum flexibility and non-linearity. It provides a single bounded sensitivity parameter $\rho \in [-1,+1]$ that is easily interpreted as the measure of non-causal association between exposure and outcome due to unobserved confounding. Moreover, our method can accommodate Bayesian inference over the ACE bounds. Thus, our method enhances an applied researcher's causal toolbox. 

\begin{figure}[t!]
\begin{center}
\begin{tikzpicture}[scale = 1.4] 
\tikzset{vertex/.style = {shape=circle,draw,minimum size=1.5em}}
\tikzset{edge/.style = {->}}
\node[vertex,dashed] (Z_A) at (0,3) {$Z_A$};
\node[vertex,dashed] (Z_Y) at (8,3) {$Z_Y$};
\node[vertex,dashed] (U_A) at (0,1.5) {$U_A$};
\node[vertex,dashed] (U_Y) at (8,1.5) {$U_Y$};
\node[vertex,dashed] (e_A) at (0,0) {$\varepsilon_A$};
\node[vertex,dashed] (e_Y) at (8,0) {$\varepsilon_Y$};
\node[vertex] (A) at (0,-1.5) {$A$};
\node[vertex] (Y) at (8,-1.5) {$Y$};
\draw[edge] (A) to node[above,sloped] {{$F_{A,Y}(A,Y)$}} (Y);
\draw[edge,dashed] (e_A) to node[right] {$t_A(\cdot)$} (A);
\draw[edge,dashed] (e_Y) to node[left] {$t_{Y|A}(\cdot)$} (Y);
\draw[-,dashed] (e_Y) to node[above,sloped] {$F_{\varepsilon_A,\varepsilon_Y}(\varepsilon_A,\varepsilon_Y) = C(F_{\varepsilon_A}(\varepsilon_A),F_{\varepsilon_Y}(\varepsilon_Y))$} (e_A);
\draw[edge,dashed] (U_A) to node[right] {$F^{-1}_{\varepsilon_A}(\cdot)$} (e_A);
\draw[edge,dashed] (U_Y) to node[left] {$F^{-1}_{\varepsilon_Y}(\cdot)$} (e_Y);
\draw[-,dashed] (U_Y) to node[above,sloped] {$F_{U_A,U_Y}(U_A,U_Y) = C(U_A,U_Y)$} (U_A);
\draw[edge,dashed] (Z_A) to node[right] {$\Phi(\cdot)$} (U_A);
\draw[edge,dashed] (Z_Y) to node[left] {$\Phi(\cdot)$} (U_Y);
\draw[-,dashed] (Z_Y) to node[above] {$F_{Z_A,Z_Y}(Z_A,Z_Y) = C(\Phi(Z_A),\Phi(Z_Y)) \approx \Phi_\rho(Z_A,Z_Y)$} (Z_A);
\draw[edge] (A) to (Y);
\end{tikzpicture} 
\end{center}
\caption{ 
Graphical representation of Equations \ref{seq:eps dist scm1}-\ref{seq:eps dist copula_confounding2}.
}
\label{sfig:copula_confounding_cgnf_rho}
\end{figure}
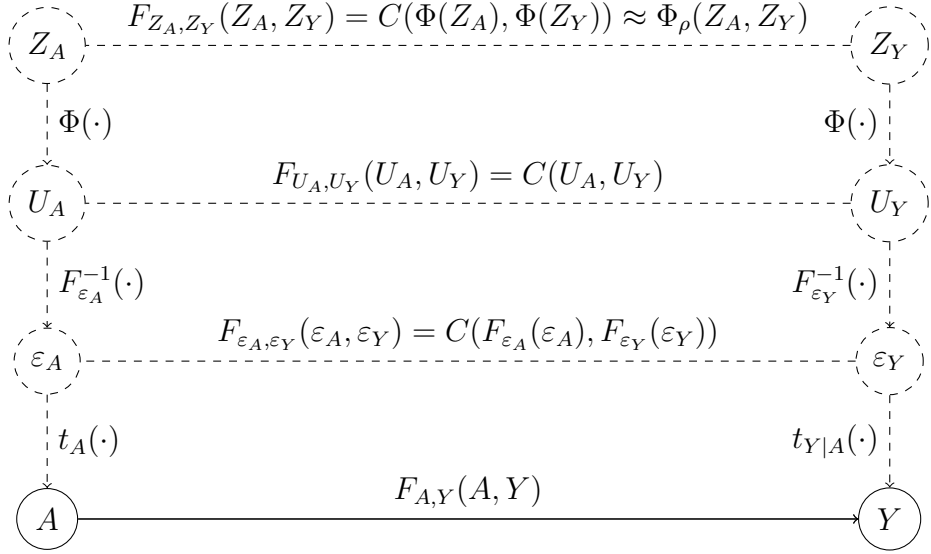

\section{Sensitivity Analysis with Copula-based Normalizing Flows}\label{sec:rho-GNF}

\subsection{Notation and Problem Definition}

Let us consider the following the structural causal model (SCM) \citep{pearl2009causality,peters2017eci}, where $A$ is the treatment (or exposure or cause) and $Y$ is the outcome (or effect), and $\varepsilon_A$ and $\varepsilon_Y$ are their respective unobserved causes such that
\begin{equation}\label{seq:eps dist scm1}
A := t_A(\varepsilon_A)  \enspace, \enspace Y := t_Y(A,\varepsilon_Y) = t_{Y|A}(\varepsilon_Y) \enspace, \enspace (\varepsilon_A, \varepsilon_Y) \sim F_{\varepsilon_A,\varepsilon_Y}(\varepsilon_A,\varepsilon_Y)
\end{equation}
where $t_A$ and $t_{Y|A}$ are arbitrary transformations, and $(\varepsilon_A, \varepsilon_Y)$ follows an arbitrary joint CDF $F_{\varepsilon_A,\varepsilon_Y}(\varepsilon_A,\varepsilon_Y)$. 
 
Using the universality of the uniform (also known as the probability integral transform) \citep{angus1994probability}, the noise variables $\varepsilon_A$ and $\varepsilon_Y$ of the SCM in Equation \ref{seq:eps dist scm1} can equivalently be written in terms of uniform variables $U_A$ and $U_Y$ in the interval $[0,1]$ resulting in the following SCM:
\begin{equation}\label{seq:eps dist copula_confounding}
A := t_A(F_{\varepsilon_A}^{-1}(U_A))  \enspace, \enspace Y := t_{Y|A}(F_{\varepsilon_Y}^{-1}(U_Y)) \enspace , \enspace (U_A, U_Y) \sim F_{U_A,U_Y}(U_A,U_Y)
\end{equation}
where $F_{\varepsilon_A}$ and $F_{\varepsilon_Y}$ respectively denote the marginal CDFs of $\varepsilon_A$ and $\varepsilon_Y$, and $(U_A, U_Y)$ follows the joint CDF $F_{U_A,U_Y}(U_A,U_Y)$ with uniform marginals in [0,1]. From the universality of the uniform, Equation \ref{seq:eps dist copula_confounding} can further be simplified in terms of $F_{A}$ and $F_{Y|A}$, which respectively denote the marginal CDFs of $A$ and $Y$ conditioned on $A$:
\begin{equation}\label{seq:eps dist copula_confounding1}
A := F_{A}^{-1}(U_A)  \enspace, \enspace Y := F_{Y|A}^{-1}(U_Y) \enspace , \enspace  (U_A, U_Y) \sim F_{U_A,U_Y}(U_A,U_Y) \enspace .
\end{equation}
Furthermore, we can represent $U_A$ and $U_Y$ in Equation \ref{seq:eps dist copula_confounding1} as transformations of standard normal variables using the standard normal CDF $\Phi$:
\begin{equation}\label{seq:eps dist copula_confounding2}
A := F_{A}^{-1}(\Phi(Z_A)) \enspace, \enspace Y := F_{Y|A}^{-1}(\Phi(Z_Y)) \enspace,\enspace  (Z_A, Z_Y) \sim F_{Z_A,Z_Y}(Z_A,Z_Y)
\end{equation}
where $(Z_A, Z_Y)$ follows the joint CDF $F_{Z_A,Z_Y}(Z_A,Z_Y)$ with standard normal marginals. A graphical representation of Equations \ref{seq:eps dist scm1}-\ref{seq:eps dist copula_confounding2} can be found in Figure \ref{sfig:copula_confounding_cgnf_rho}.

Equations \ref{seq:eps dist scm1}-\ref{seq:eps dist copula_confounding2} represent observationally and interventionally equivalent SCMs, but with different unobserved noises and corresponding joint CDFs \citep{pearl2009causality,mooij2016distinguishing}. Observational equivalence means that the SCMs yield the same distribution $F_{A,Y}(A,Y)$. Interventional equivalence means that the SCMs yield the same interventional distributions $F_{Y}(Y|do(a))$ and $F_{A}(A|do(y))$. 
Since these noises are unobserved, the interventional distribution of interest $F_{Y}(Y|do(a))$ is not identifiable from any of the SCMs, without further assumptions \citep{pearl2009causality}. To overcome this problem, we propose in the subsequent sections to use a bivariate Gaussian copula to model the noise distributions (i.e., confounding strength) so that the interventional distribution of interest can parametrically be estimated using deep-neural-network-inspired normalizing flows trained on observational data.

\subsection{Representing Confounding with a Gaussian Copula}

A copula is a multivariate distribution function defined on the unit hypercube with uniform marginals \citep{nelsen2007introduction}. As the name suggests, a copula ties or links or couples a multidimensional joint distribution to its marginals. Therefore, by Sklar's theorem \citep{nelsen2007introduction}, we have that the bivariate joint CDF $F_{U_A,U_Y}(U_A,U_Y)$ in Equations \ref{seq:eps dist copula_confounding} and \ref{seq:eps dist copula_confounding1} with uniform marginals in [0,1] can be represented using some bivariate copula $C(U_A,U_Y)$:
\begin{align}
F_{U_A,U_Y}(U_A,U_Y) &= C(U_A,U_Y)\label{seq:sklarstheorem}\\
F_{\varepsilon_A,\varepsilon_Y}(\varepsilon_A,\varepsilon_Y) &= C(F_{\varepsilon_A}(\varepsilon_A),F_{\varepsilon_Y}(\varepsilon_Y))\label{seq:sklarstheorem1}\\
F_{Z_A,Z_Y}(Z_A,Z_Y) &= C(\Phi(Z_A),\Phi(Z_Y))\label{seq:sklarstheorem2}\enspace.
\end{align}
Equations \ref{seq:sklarstheorem1} and \ref{seq:sklarstheorem2} follow from the scale-invariance property of the copula $C(U_A,U_Y)$ to the strictly increasing transformations $F_{\varepsilon_A}$, $F_{\varepsilon_Y}$, and $\Phi$ \citep{nelsen2007introduction}. The copula essentially models the confounding effect (i.e., non-causal back-door association) between exposure and outcome in Equations \ref{seq:eps dist scm1}-\ref{seq:eps dist copula_confounding2}. See also Figure \ref{sfig:copula_confounding_cgnf_rho}. This may be quantified by measures of association between $U_A$ and $U_Y$ such as Spearman correlation $\rho_S$ \citep{spearman2010proof} or Kendall correlation $\tau_K$ \citep{kendall1938new}. By the scale-invariance property of $\rho_S$ and $\tau_K$ to the strictly increasing transformations $F_{\varepsilon_A}$, $F_{\varepsilon_Y}$, and $\Phi$, these measures of association between $U_A$ and $U_Y$ are the same as between $\varepsilon_A$ and $\varepsilon_Y$, and between $Z_A$ and $Z_Y$.

As discussed above, the interventional distribution of interest $F_{Y}(Y|do(a))$ is identifiable when the copula $C(U_A,U_Y)$ is known so as to adjust for the non-causal back-door path between $A$ and $Y$ \citep{pearl2009causality}. However, the copula, although uniquely exists \citep{nelsen2007introduction}, remains unknown and cannot be estimated from observational data as the noises are unobserved. Since the copula is unknown and unlearnable, it is inevitable to make assumptions about it to achieve causal effect identification. Specifically, the copula may be chosen from any of the vast families in the literature such as Archimedean, elliptical, or empirical copulas \citep{nelsen2007introduction, salvadori2007extremes, durante2016principles}. For instance, the Gaussian copula is one of the most studied elliptical copulas and it has been widely used in the fields of quantitative finance \citep{cherubini2004copula, salmon2009recipe, mackenzie2014formula}, hydrology research \citep{renard2007use, zhang2019copulas}, logistics \citep{kumar2019copula}, astronomy \citep{takeuchi2010constructing}, and others \citep{nelsen2007introduction, salvadori2007extremes, durante2016principles}. Recent works such as \citep{zheng2021copula, zheng2022sensitivity} have proposed sensitivity analysis with the Gaussian copula, but without the use of normalizing flows. In this work, we thus approximate the unknown copula $C(U_A,U_Y)$ with the Gaussian copula. That is,
\[
C(U_A,U_Y) \approx \Phi_\rho(\Phi^{-1}(U_A),\Phi^{-1}(U_Y))
\]
which yields
\[
F_{Z_A,Z_Y}(Z_A,Z_Y) = C(\Phi(Z_A),\Phi(Z_Y)) \approx \Phi_\rho(Z_A,Z_Y)
\]
where $\rho \in [-1,+1]$ is the Pearson's correlation between $Z_A$ and $Z_Y$. See Figure \ref{sfig:copula_confounding_cgnf_rho}. As we will see, the latter expression fits really well into normalizing flows for modeling the SCM in Equation \ref{seq:eps dist copula_confounding2}. It is also worth noticing that the Gaussian copula parameter $\rho$ approximately denotes the confounding strength $\rho_{S}$, since $\rho=2\sin(\pi\rho_{S}/6)$ \citep{kruskal1958ordinal, meyer2013bivariate}.

\subsection{ACE Estimation with Normalizing Flows}\label{sec:MC}

Since $F_{\varepsilon_A}$, $F_{\varepsilon_Y}$, and $\Phi$ are strictly increasing functions, we can rewrite Equation \ref{seq:eps dist copula_confounding2} as
\begin{equation}\label{eq:rho-GNF}
A := T_A^{-1}(Z_A)  \enspace, \enspace Y := T_{Y|A}^{-1}(Z_Y) \enspace,\enspace (Z_A, Z_Y) \sim \Phi_\rho(Z_A,Z_Y)
\end{equation}
where $T_A$ and $T_{Y|A}$ are strictly monotonic, and thus invertible, transformations of the arbitrarily distributed observed random vector $(A,Y)$ into the normally distributed random vector $(Z_A,Z_Y)$. Such transformations are aptly called normalizing flow, and they are typically modeled as deep neural networks \citep{Papamakarios2017MAF, papamakarios2021NF_pmi, kobyzev2020NF}. Specifically, we use unconstrained monotonic neural networks for the transformations \citep{wehenkel2019UMNN,balgi2022cgnf}, and the graphical conditioner neural network for conditioning the transformation of $Y$ on its parent $A$ \citep{wehenkel2020GNF,balgi2022cgnf}. Such a normalizing flow is able to universally model any arbitrary data distribution \citep{Huang2018NAF}. We henceforth refer to it as $\rho$-GNF.

As any normalizing flow, our $\rho$-GNF is trained by maximizing the $\log$-likelihood of a observational dataset for a fixed $\rho$ value. More specifically, let $X=(A,Y)$, $Z=(Z_A,Z_Y)$, and $T(Z_A,Z_Y;\theta)=(T_A(Z_A;\theta_A),T_{Y|A}(Z_Y;\theta_Y))$ where we explicitly represent the parameters of the deep neural networks as $\theta = (\theta_A,\theta_Y)$. Let $\{X^\ell\}^{N}_{\ell{ = }1}$ denote an observational dataset. Then, the $\log$-likelihood can be expressed as follows by a change of variables \citep{Papamakarios2017MAF, papamakarios2021NF_pmi, kobyzev2020NF}:
\[
\mathcal{LL(\theta)}{ = }\sum_{\ell{ = }1}^{N}\mathrm{log} \left(f_X(X^\ell;\theta)\right)
\]
where
\[ 
f_X(X^\ell;\theta){ = }f_Z\left(T^{-1}(X^\ell;\theta)\right) \cdot \left|\mathrm{det}\left( J_{T^{-1}(X^\ell;\theta)}(X^\ell)\right)\right|.
\]
Recall that $T$ is invertible by construction of normalizing flows. For the same reason, the determinant of the Jacobian $\mathrm{det}\left( J_{T^{-1}(X^\ell;\theta)}(X^\ell)\right)$ can be computed efficiently \citep{Papamakarios2017MAF, papamakarios2021NF_pmi, kobyzev2020NF}. Under our Gaussian copula assumption, $f_Z(Z^\ell)$ is simply a bivariate normal density function. Therefore, our $\rho$-GNF can be trained efficiently.

Our main objective in this work is to estimate the average causal effect (ACE) as a function of $\rho$, which can be expressed as
\[
ACE_{\rho} = E[Y|do(A:=1)] - E[Y|do(A:=0)] =E[Y_1] - E[Y_0]
\]
where $Y_a$ denotes the potential outcome under the intervention $A{:=}a$. In particular, we use Morte-Carlo estimation by drawing samples from the interventional distribution $F_{Y}(Y|do(a))$ after having trained the $\rho$-GNF for a fixed $\rho$ value. More concretely, we follow the following three steps:
\begin{enumerate}
\item $Z_Y^\ell{=}T_{Y|A^\ell}({Y}^\ell;\theta_Y) \enspace \text{for} \enspace   \ell = 1,\ldots,N$.
\\
\item $Y^\ell_a{=}T^{-1}_{Y|a}(Z^\ell_Y;\theta_Y) \enspace  \text{for} \enspace   \ell = 1,\ldots,N$.
\\
\item $ACE_{\rho}=E[Y_1]{-}E[Y_0]{\approx}\frac{\sum^{N}_{\ell{=}1} Y^{\ell}_1}{N} - \frac{\sum^{N}_{\ell{=}1} Y^{\ell}_0}{N}$.
\end{enumerate}
The first step recovers the unobserved noise values for the observations in the training data, the second step computes the potential outcomes for the recovered noises, and the third step produces the Monte-Carlo estimates. The first step is possible due to the invertibility of normalizing flows. The first step can also be replaced by drawing $N$ samples from $\Phi_\rho(Z_A,Z_Y)$.

\subsection{Sensitivity Analysis with $\rho$-GNF}

As mentioned before, our main aim in this work is to estimate the ACE as a function of $\rho$, so as to determine how sensitivity the ACE is to different degrees of confounding. We have shown how to do it with the $\rho$-GNF. The result can be summarized in a sensitivity plot (ACE against $\rho)$ that we aptly refer to as the $\rho_{curve}$. Note that, unlike in other sensitivity analysis methods, our sensitivity parameter $\rho$ is conveniently bounded and interpretable. We illustrate this with examples in Section \ref{sec:experiments}.

One of the most popular sensitivity analysis methods is based on the so-called E-value, which is defined as the minimum strength of association on the risk ratio scale that an unmeasured confounder would need to have with both treatment and outcome to fully explain away the observed treatment–outcome association, conditional on the measured covariates \citep{vanderweele2017sensitivity}. A large E-value thus implies that considerable unmeasured confounding would be needed to nullify the causal effect. Likewise, a small E-value implies that little unmeasured confounding would be needed to nullify the causal effect. Similar to the E-value, we propose the $\rho_{value}$ which represents the Gaussian copula parameter value that explains away the observed association between treatment and outcome. In other words, setting $\rho = \rho_{value}$ results in $ACE_{\rho} = 0$. Let $\rho_{S_{Obs}}$ denote the Spearman correlation between the treatment $A$ and the outcome $Y$ in the observational data at hand. When $A$ has no causal effect on $Y$, we have that $\rho_S(Z_A,Z_Y) = \rho_{S_{Obs}}$ by the scale-invariance property of Spearman correlation to the strictly increasing transformations $T^{-1}_{A}$ and $T^{-1}_{Y|A}=T^{-1}_{Y}$. Therefore, $\rho_{value}{=}2sin\left( \pi \rho_{S_{Obs}}/6\right)$. 

Moreover, the $\rho_{curve}$ and $\rho_{value}$ may help the analyst to determine the sign of the ACE (i.e., whether the treatment is harmful or beneficial), which is arguably the most important part of sensitivity analysis. Specifically, suppose the analyst hypothesizes a measure of confounding in the interval $[\rho_{min},\rho_{max}]$. Thus, the $\rho_{curve}$ enables her to bound the ACE to the narrower interval $[ACE_{\rho_{max}},ACE_{\rho_{min}}]$, which may in turn help her to determine the sign of the ACE: If $\rho_{min} > \rho_{value}$ (or $\rho_{max} < \rho_{value}$) then she may conclude that the ACE is negative (or positive). We illustrate this with examples in Section \ref{sec:experiments}.

\subsection{On the Gaussian Copula Assumption}\label{sec:assumption}

We close this section with a discussion on the Gaussian copula assumption. In principle, any copula may be assumed in place of the Gaussian copula in Equations \ref{seq:sklarstheorem}-\ref{seq:sklarstheorem2}. However, the Gaussian copula assumption makes it possible to seamlessly integrate these equations into normalizing flows and thus produce our $\rho$-GNF, which may be seen as a generalization of the ordinary normalizing flow \citep{Papamakarios2017MAF, papamakarios2021NF_pmi, kobyzev2020NF}. Specifically, the ordinary normalizing flow corresponds to $\rho$-GNF with $\rho=0$. The fact that $Z_A$ and $Z_Y$ are still normally distributed makes the $\rho$-GNF retain one of the most salient features of the ordinary normalizing flow, namely the efficient computation of the $\log$-likelihood of the training dataset which enables computationally efficient training of the $\rho$-GNF. Furthermore, the Gaussian copula assumption enables efficient sampling of $(Z_A, Z_Y)$ for Monte-Carlo estimation of the ACE, thus enabling computationally efficient causal inference. Finally, the Gaussian copula assumption provides a single bounded and interpretable sensitivity parameter $\rho \in [-1,+1]$ for sensitivity analysis that can be used to control/model/adjust the back-door non-causal association under which the ACE is identifiable. Thus, enabling simple and efficient sensitivity analysis. Our subsequent experiments and results show that the Gaussian copula assumption works well empirically.

\section{Bayesian Sensitivity Analysis with $\rho$-GNF}\label{sec:Bayesian}

In this section, we extend our method for sensitivity analysis by approaching it from a Bayesian perspective. Defining a prior distribution over the sensitivity parameter (i.e., the $\p$ parameter of the Gaussian copula) makes it possible to derive the posterior distribution over the ACE, which makes it possible to calculate credible intervals.

Let $Q$ be the causal quantity the analyst is interested in. In this work, $Q$ is the ACE but our method actually applies to any causal quantity computable from the $\rho$-GNF (e.g., the expected potential outcome under some treatment). Then, calculating $Q$ with the $\rho$-GNF will result in different values of $Q$ for different values of $\p$. Consequently, we can define $Q$ as a function $h$ of $\p$ such that $Q = h(\p)$. However, we do not have any analytic expression of the function $h$, but we can evaluate it for a discrete set of points $\p \in \rhoSet = \{\p_1, \p_2, \ldots, \p_n\}$ by training the $\rho$-GNF for those values of $\p$ and evaluating $Q$. Next, we present two approaches for estimating $h$ from the evaluation points $\rhoSet$, to be used to produce the posterior distribution of $Q$.

\paragraph{Continuous function approximation} The first approach that we present is to make use of a change of variable to express the posterior distribution over $Q$ as
\begin{equation}\label{eq:continuous-q}
f_{Q}(Q=q) = f_{\p}\left(h^{-1}(q) \right) \cdot \left| \frac{dh^{-1}(q)}{dq} \right|
\end{equation}
where $f_{\p}(\p)$ is a prior distribution over $\p$, and then approximate $h$ as a continuous function $\Tilde{h}$ by using some method to interpolate the values $\left[-1, 1 \right] \setminus \rhoSet$ for which $h$ is not evaluated. For example, $\Tilde{h}$ could be a piecewise linear function with $n$ degrees of freedom such that $\Tilde{h}(\p) = h(\p)$ for $\p \in \rhoSet$.

Unfortunately, this approach comes with several disadvantages. First, it is only possible if the inverse of $h$ exists. This might not always be true in reality. Secondly, even if the inverse of $h$ exists, it may be too badly approximated. For instance, let us suppose that $h(\p) = \p$. Due to the stochastic training process of the $\rho$-GNF, we may get some bias for our evaluation points so that $\Tilde{h}(\p) = h(\p) + \epsilon$ where $\epsilon$ represents the bias. Then, for the derivative in point $\p_0 \in \rhoSet$, we have that
\begin{equation*}
    \frac{d\Tilde{h}(\p_0)}{d\p} \approx \lim_{\p_0 - \Tilde{\p} \to 0} \frac{\Tilde{h}\left( \p_0 \right) - \Tilde{h}\left( \Tilde{\p} \right)}{\p_0 - \Tilde{\p}} = \lim_{\p_0 - \Tilde{\p} \to 0} \frac{\p_0 + \epsilon_0 - \Tilde{\p} - \Tilde{\epsilon}}{\p_0 - \Tilde{\p}} \in \{ -\infty, 1, +\infty \} \enspace .
\end{equation*}
Observe that the derivative is $-\infty$ or $+\infty$ almost surely. It is easy to verify that the derivative of the inverse of $\Tilde{h}$ suffers from the same limitation. This is problematic when plugged into Equation \ref{eq:continuous-q}.

\paragraph{Discrete function approximation} We present another approach for estimating $h$ that involves representing it as a discrete function. Let the discrete approximation of $h$ be denoted $\Tilde{h}: \rhoSet \rightarrow \QSet \subset \R$ with inverse $\Tilde{h}^{-1}: \QSet \rightarrow \rhoSet$. We furthermore redefine $\p$ as a discrete random variable taking values in $\rhoSet$. Without loss of generality, we assume that $\p_i < \p_{i + 1}$ for $i=1,\ldots, n-1$ and define
\begin{align*}
    P(\p_i) &= F_{\p}\left(\frac{1}{2} \cdot (\p_i + \p_{i+1})\right) - F_{\p}\left(\frac{1}{2} \cdot (\p_i + \p_{i-1})\right) \text{ for } i=2,\ldots, n-1 \\
    P(\p_1) &= F_{\p}\left(\frac{1}{2} \cdot (\p_1 + \p_{2})\right) \\
    P(\p_n) &= 1 - F_{\p}\left(\frac{1}{2} \cdot (\p_n + \p_{n-1})\right)
\end{align*}
where $F_\p$ is the CDF of $\p$ corresponding to the prior distribution over $\rho$. Then, we can derive $P(Q=q)$ for $q \in \QSet$ as
\begin{equation}\label{eq:bayesian-density-estimation}
P(Q=q) = P\left(\p \in \Tilde{h}^{-1}(q) \right) = \sum_{r \in \Tilde{h}^{-1}(q)} P\left(\p = r  \right).
\end{equation}

Note that because we assume that $\p$ is discrete, $Q$ becomes a discrete random variable as well. We can transform the discrete distribution of $Q$ back to a continuous distribution by using kernel smoothing \citep{Wand1994}:
\begin{equation}\label{eq:kernel}
    f_Q(Q=q) = \sum_{\Tilde{q} \in \QSet} K(q - \Tilde{q}) \cdot P(Q=\Tilde{q})
\end{equation}
where $K$ is a kernel function.

\subsection{On the Choice of the Prior Distribution of $\rho$}\label{sec:prior}

In this section, we discuss how a suitable prior distribution for $\p$ can be chosen to reflect expert knowledge about the data generating process. There are two main reasons for the noise variables $\varepsilon_A$ and $\varepsilon_Y$ to be dependent in the SCM in Equation \ref{seq:eps dist scm1} and Figure \ref{sfig:copula_confounding_cgnf_rho}: Selection bias and hidden confounding \citep{pearl2009causality}. Selection bias occurs when conditioning on some common effect of the noise variables. Hidden confounding occurs when the noise variables have some common cause. As the latter is the most commonly studied scenario in the literature \citep{pearl2009causality}, we focus on it hereinafter. We start by discussing what different values of $\p$ imply about the hidden confounder. Based on these observations, we then make suggestions on how to do the reverse, i.e., how to derive a prior function over $\p$ that matches the existing expert knowledge about the hidden confounder.

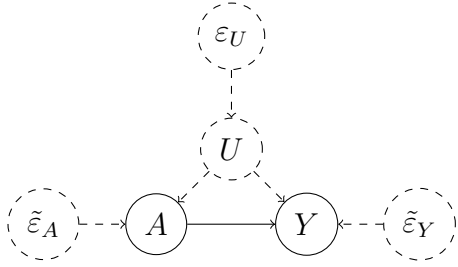
\begin{figure}[t!]
    \centering
    \begin{tikzpicture}
        \node[draw, circle, dashed] (eps_U) at (1,2.5) {$\varepsilon_U$};
        \node[draw, circle, dashed] (U) at (1,1) {$U$};
        \node[draw, circle, dashed] (eps_A) at (-1.5,0) {$\Tilde{\varepsilon}_A$};
        \node[draw, circle] (A) at (0,0) {$A$};
        \node[draw, circle, dashed] (eps_Y) at (3.5,0) {$\Tilde{\varepsilon}_Y$};
        \node[draw, circle] (Y) at (2,0) {$Y$};
        
        \draw[->, dashed] (eps_U) -- (U);
        \draw[->, dashed] (U) -- (A);
        \draw[->, dashed] (eps_A) -- (A);
        \draw[->, dashed] (U) -- (Y);
        \draw[->, dashed] (eps_Y) -- (Y);
        \draw[->] (A) -- (Y);
    \end{tikzpicture}
    \caption{Causal graph of the SCM in Proposition \ref{prop:equivalent-scms}.}
    \label{fig:dag-hidden-confounder}
\end{figure}

\begin{proposition}\label{prop:equivalent-scms}
The SCM in Equation \ref{eq:rho-GNF} and depicted in Figure \ref{sfig:copula_confounding_cgnf_rho} can be rewritten as an observationally and interventionally equivalent SCM with a hidden confounder as depicted in Figure \ref{fig:dag-hidden-confounder} by expressing it in the form of the following SCM:
\begin{align*}
U &:= \varepsilon_U \\
A &:= \Tilde{t}_A\left(U,  \Tilde{\varepsilon}_A\right) = T_A^{-1}\left(\frac{1}{\sqrt{\gamma^2 + \delta^2}} \left( \gamma U + \delta \Tilde{\varepsilon}_A \right) \right) \\
Y &:= \Tilde{t}_{Y}\left(U, A, \Tilde{\varepsilon}_Y\right) = T_{Y|A}^{-1}\left(A, \lambda \p U + \tau \sqrt{1-\p^2} \Tilde{\varepsilon}_Y \right) \\
        \varepsilon_U &, \Tilde{\varepsilon}_A, \Tilde{\varepsilon}_Y \simiid \Normal{0}{1} \\
        \gamma &\in \R \setminus \{0\}\\
        \delta &\in \left[-\sqrt{\frac{(1 - \p^2) \gamma^2}{\p^2}},\ \sqrt{\frac{(1 - \p^2) \gamma^2}{\p^2}} \right] \\
        \lambda &= \frac{\sqrt{\gamma^2 + \delta^2}}{\gamma} \\
        \tau &= \sqrt{\frac{1}{1 - \p^2} - \frac{\gamma^2 + \delta^2}{\gamma^2} \cdot \frac{\p^2}{1 - \p^2}} \enspace .
\end{align*}
\end{proposition}

The proof of the proposition can be found in \ref{proposition}. The main idea is to rewrite the Gaussian noise random vector $(Z_A, Z_Y)$ as a function of three independent Gaussian noise random variables $\Tilde{\varepsilon}_A, \Tilde{\varepsilon}_Y$ and $\varepsilon_U$.

Let the influence of the random variable $U$ on another variable $X$ be defined as $\partialDer{X}{U}$.

\begin{theorem}\label{thm:sign-of-rho}
The influence of $U$ on $A$ and the influence of $U$ on $Y$ in the SCM in Proposition \ref{prop:equivalent-scms} have the same sign if and only if $\p > 0$. The specific sign depends on the choice of $\gamma$.
\end{theorem}

The proof of the theorem can be found in \ref{theorem}.

\begin{corollary}
For the SCM in Proposition \ref{prop:equivalent-scms}, we cannot determine the sign of the effect of $U$ on $A$ or of the effect of $U$ on $Y$,  because the sign changes with the value of $\gamma$ and any value of $\gamma \in \R \setminus 0$ yields the same observational and interventional distribution.
\end{corollary}

\begin{figure}[t!]
  \begin{center}
    \begin{subfigure}[t]{0.45\textwidth}
      \centering
      \includegraphics[width=1\textwidth]{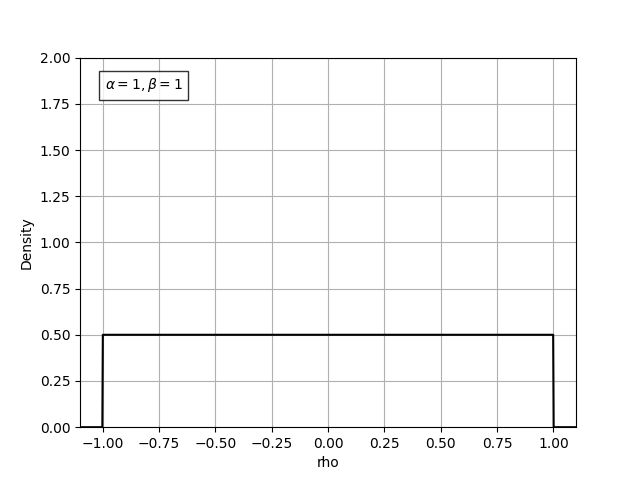}
      \caption{Uniform prior.}
      \label{fig:rho-prior1}
    \end{subfigure}
    \quad
    \begin{subfigure}[t]{0.45\textwidth}
      \centering
      \includegraphics[width=1\textwidth]{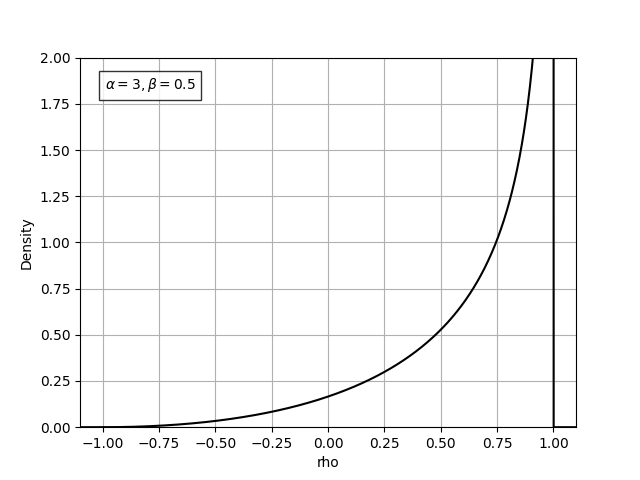}
      \caption{Influence has same sign.}
      \label{fig:rho-prior2}
    \end{subfigure}
    \quad
    \begin{subfigure}[t]{0.45\textwidth}
      \centering
      \includegraphics[width=1\textwidth]{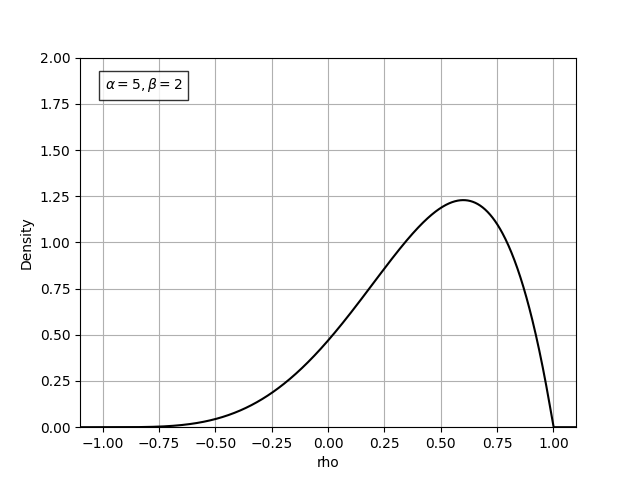}
      \caption{Influence has same sign.}
      \label{fig:rho-prior3}
    \end{subfigure}
    \quad
    \begin{subfigure}[t]{0.45\textwidth}
      \centering
      \includegraphics[width=1\textwidth]{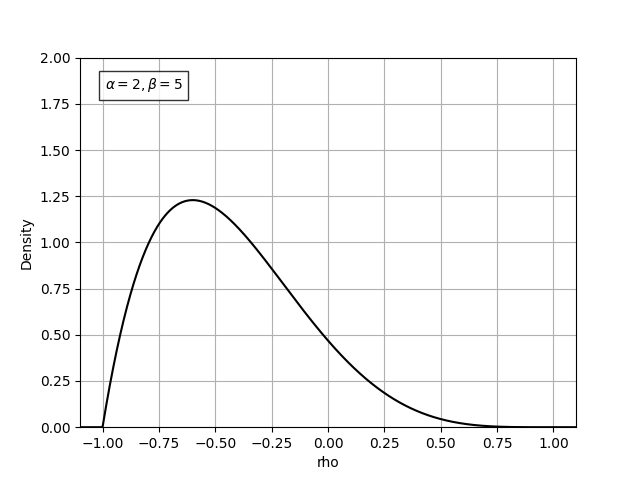}
      \caption{Influence has opposite sign.}
      \label{fig:rho-prior4}
    \end{subfigure}
    \quad
    \begin{subfigure}[t]{0.45\textwidth}
      \centering
      \includegraphics[width=1\textwidth]{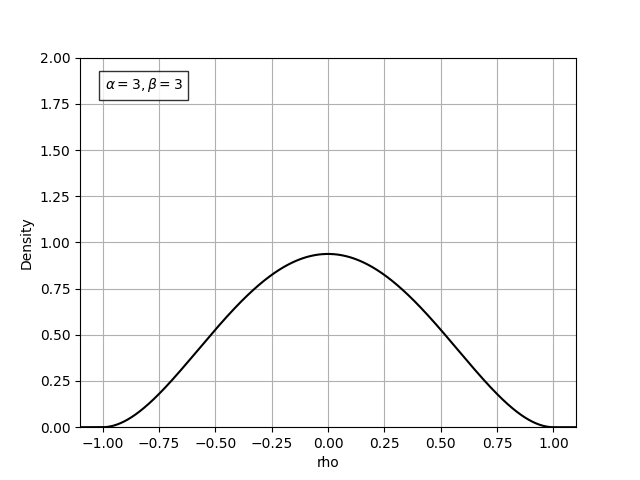}
      \caption{No hidden confounding.}
      \label{fig:rho-prior6}
    \end{subfigure}
    \quad
    \begin{subfigure}[t]{0.45\textwidth}
      \centering
      \includegraphics[width=1\textwidth]{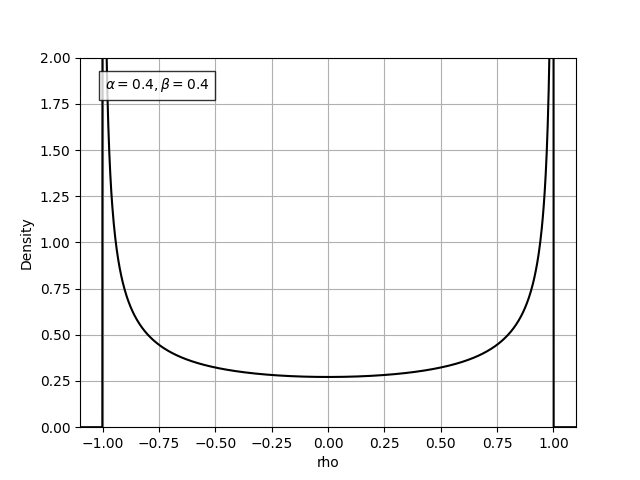}
      \caption{Hidden confounding with unknown influence.}
      \label{fig:rho-prior5}
    \end{subfigure}
  \end{center}
  \caption{Scaled and shifted beta distribution with different parameters $\alpha$ and $\beta$ suitable for using as $\p$ prior. The subcaptions describe the beliefs about hidden confounding that are encoded in the distributions. The term "influence" in the subcaption refers to the influence of the hidden confounder on the treatment and outcome variables.}
  \label{fig:rho-priors}
\end{figure}

Based on the above statements, we now make suggestions for a suitable $\p$ prior based on the believed influence of the hidden confounder on the treatment and outcome variables. We know that $\p \in \left[-1, +1\right]$, and thus we need to choose a distribution whose PDF is zero elsewhere. Suitable prior functions are therefore for example a scaled and shifted beta distribution $f_\p(\p) = \frac{1}{2} f_{\text{Beta}}\left(\frac{1}{2} (\p + 1) \right)$ or a truncated normal distribution that is truncated between $-1$ and $+1$. The choice of $\p$ depends on the believed influence of the hidden confounder on the treatment and outcome variables. If one has no prior beliefs about it, a uniform prior between -1 and +1 is the natural choice. This equates to the beta distribution with parameters $\alpha = \beta = 1$ as shown in Figure \ref{fig:rho-prior1}.

From Theorem \ref{thm:sign-of-rho} we know that, believing that there exists a hidden confounder that has either a positive or a negative influence on both the treatment and the outcome, one should choose a prior distribution that puts more weight on positive values. This could for example be a truncated normal distribution with positive mean or a beta distribution with $\alpha > \beta$ as shown in Figures \ref{fig:rho-prior2} or \ref{fig:rho-prior3}. If one believes that the influence of the confounder on the treatment has the opposite sign compared to the influence of the confounder on the outcome, one should choose a prior that puts more weight on negative values of $\p$, like a beta distribution with $\alpha < \beta$ as shown in Figure \ref{fig:rho-prior4}.

In the case where one suspects that there is no confounding, one can choose a prior with mean zero (e.g., beta distribution with $1 < \alpha = \beta$ as shown in Figure \ref{fig:rho-prior6}). In the opposite case, where one suspects confounding but does not have any prior beliefs about the sign of the influence of the confounder on treatment and outcome, a prior that puts weight on values close to -1 and +1 should be chosen (e.g., beta distribution with $0 < \alpha = \beta < 1$ shown in Figure \ref{fig:rho-prior5}). We illustrate these scenarios with examples in the next section.

\section{Experiments}\label{sec:experiments}

We present results for three different experimental settings: Simulated data with continuous outcome in Section \ref{sec:experiments simcont}, simulated data with binary outcome in Section \ref{sec:experiments simbin}, and real-world data with categorical outcome in Section \ref{sec:real}. We present experiments for the $\rho$-GNF as well as for the Bayesian $\rho$-GNF.

Our $\rho$-GNF is implemented in PyTorch\footnote{The $\rho$-GNF code is available at \url{https://github.com/sobalgi/rhoGNF}.} by adapting the baseline code of the unconstrained monotonic neural networks \citep{wehenkel2019UMNN}\footnote{Code at \url{https://github.com/AWehenkel/Graphical-Normalizing-Flows}.} and the graphical normalizing flows \citep{wehenkel2020GNF}\footnote{Code at \url{https://github.com/AWehenkel/UMNN}.}. As normalizing flows are developed for continuous variables, we use the Gaussian dequantization trick from the causal graphical normalizing flow to model discrete variables into $\rho$-GNF \citep{balgi2022cgnf}. 

The $\rho_{curve}$ for a given observational dataset is obtained by training $\rho$-GNFs for $\rho = -0.99, -0.8, -0.6, -0.4, -0.2, 0, 0.2, 0.4, 0.6, 0.8, 0.99$. We estimate the respective $ACE_{\rho}$ as described in Section \ref{sec:MC}. Our empirical ACE bounds are obtained as the infimum and supremum of the $\rho_{curve}$, i.e., $\inf[ACE_{\rho}] \leq ACE_{true} \leq \sup[ACE_{\rho}]$. We also identify the $\rho_{value}$ that explains away the causal association.

All experiments with the Bayesian $\rho$-GNF approach are run with values $\rho = -0.99, -0.95, -0.9, -0.85, -0.8, -0.75, \ldots, 0.75, 0.8, 0.85, 0.9, 0.95, 0.99$. In the experiments, $Q$ is either the ACE or an expected potential outcome. Equations \ref{eq:bayesian-density-estimation} and \ref{eq:kernel} are used to estimate the distribution of the ACE and the expected potential outcome. As kernel function, we use a Gaussian kernel with variance $\frac{1}{16} \var{\QSet}$ whenever plotting the density of $Q$.

\begin{table*}[t!]
  \centering
  \begin{tabular}{l|ccc|cccc}
    \hline
$SCM_{\alpha,\beta,\delta}$ & $\alpha$  & $\beta$ & $\delta$ & $\rho_{P_{Obs}}$ & $\rho_{true}$  & $ACE_{true}$ & $\rho_{value}$\\
\hline
$SCM_{1}$ &  0.2 & -0.6 & 0.72 & -0.55 & -0.71  & 0.2 & -0.55\\
$SCM_{2}$ &  0.0 & -0.4 & 0.52 & -0.55 & -0.55 & 0.0  & -0.55\\
$SCM_{3}$ &  -0.2 & -0.2 & 0.40 & -0.55 & -0.32  & -0.2  & -0.55\\
\hline
$SCM_{4}$ &  0.2 & 0.2 & 0.40 & 0.55 & 0.32  & 0.2  & 0.55\\
$SCM_{5}$ &  0.0 & 0.4 & 0.52 & 0.55 & 0.55 & 0.0 & 0.55\\
$SCM_{6}$ &  -0.2 & 0.6 & 0.72 & 0.55 & 0.71  & -0.2 & 0.55\\
    \hline
  \end{tabular}
  \caption{Six observationally and/or interventionally non-equivalent SCMs with different mixtures of total observed association ($\rho_{P_{Obs}}$), non-causal association ($\rho_{true}$), causal association ($ACE_{true}$), and $\rho_{value}$. Note that $ACE_{true}{=}\alpha$.} 
  \label{tbl: Jose toy SCM}
\end{table*}

\subsection{Experiments with Continuous Outcome}\label{sec:experiments simcont}

In our first set of simulated experiments, we consider the SCM with continuous treatment $A$ and outcome $Y$ in the work \citep{hoover2006causality}. This is a well-studied SCM from economics and econometrics, and is defined as follows.
\begin{equation}\label{eq:eco}
A {:=}\varepsilon_A   \enspace, \enspace\enspace Y{:=}\alpha A{+}\varepsilon_Y \enspace , \vectwod{\varepsilon_A}{\varepsilon_Y} 
        \sim
        \Normal{\vectwod{0}{0}}{\begin{pmatrix}
            1 & \beta \\
            \beta & \delta
        \end{pmatrix}}
\enspace.
\end{equation}
For given $\alpha,\beta$ and $\delta$ values, we have that the total Pearson's correlation between $A$ and $Y$ is $\rho_{P_{Obs}}{=}\frac{\sigma_{A,Y}}{\sigma_{A}\sigma_{Y}}$, where $\sigma^2_A{=}1$ and  $\sigma^2_Y{=}\alpha^2{+}\delta{+}2\alpha\beta$ and $\sigma_{A,Y}{=}\alpha{+}\beta$. The Pearson's correlation due to the non-causal path is $\rho_{true}{=}\frac{\sigma_{\varepsilon_A,\varepsilon_Y}}{\sigma_{\varepsilon_A}\sigma_{\varepsilon_{Y}}}=\beta/\delta$, while the causal one is $ACE_{true}{=}\alpha$. These values are used for verification purposes. See Table \ref{tbl: Jose toy SCM} for the $\alpha,\beta$ and $\delta$ values considered in our experiments.

\begin{figure}[t!]
  \begin{center}
    \begin{subfigure}{1\textwidth}
      \centering
      \includegraphics[width=.6\textwidth]{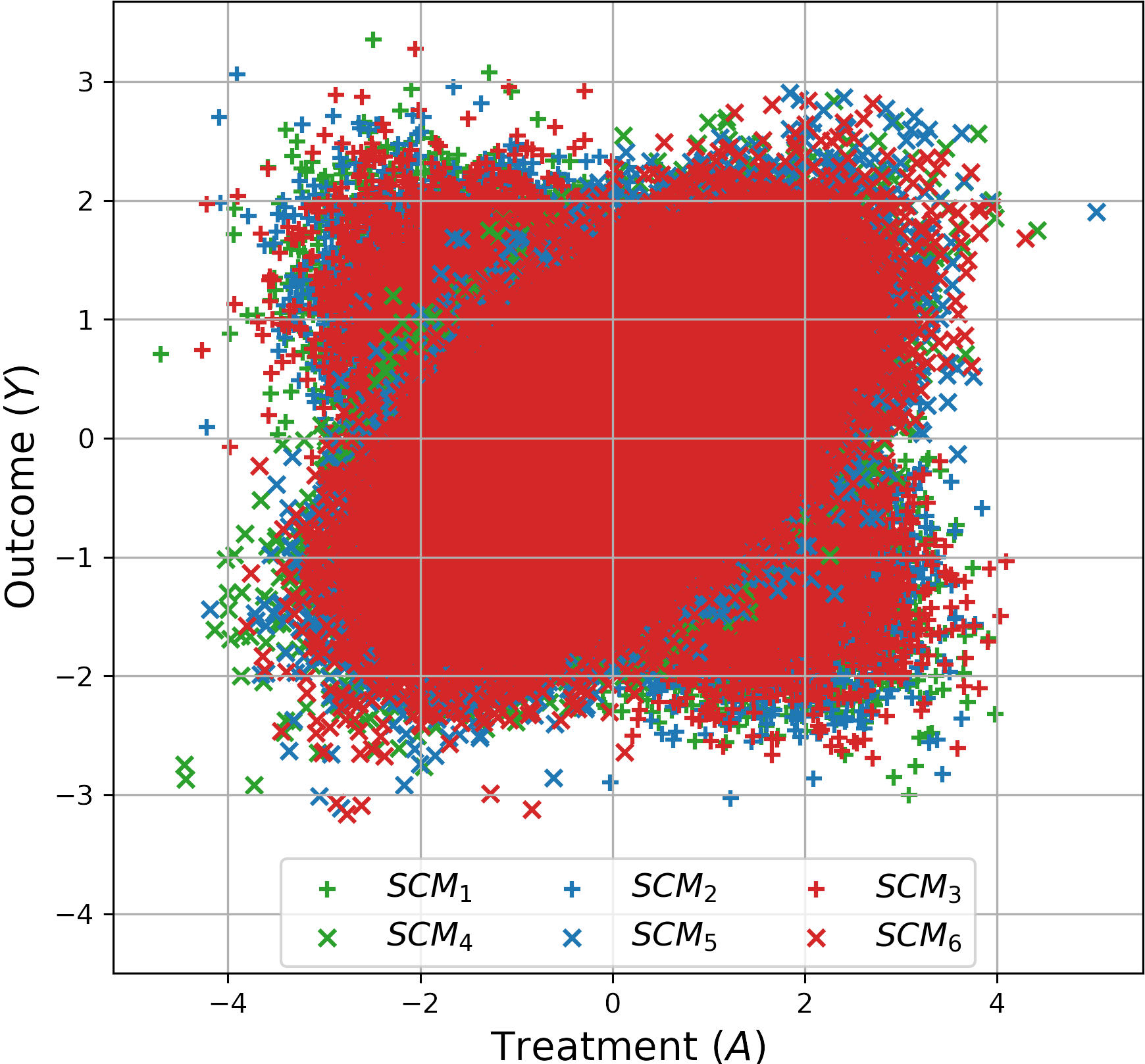}
      \caption{Scatter plots of the observational datasets sampled from the six SCMs in Table \ref{tbl: Jose toy SCM}.}
      \label{sfig:Jose_HVR_SCMs_050000_rho_curves_b}
    \end{subfigure}
    \begin{subfigure}{1\textwidth}
      \centering
      \vspace{1cm}
      \includegraphics[width=.8\textwidth]{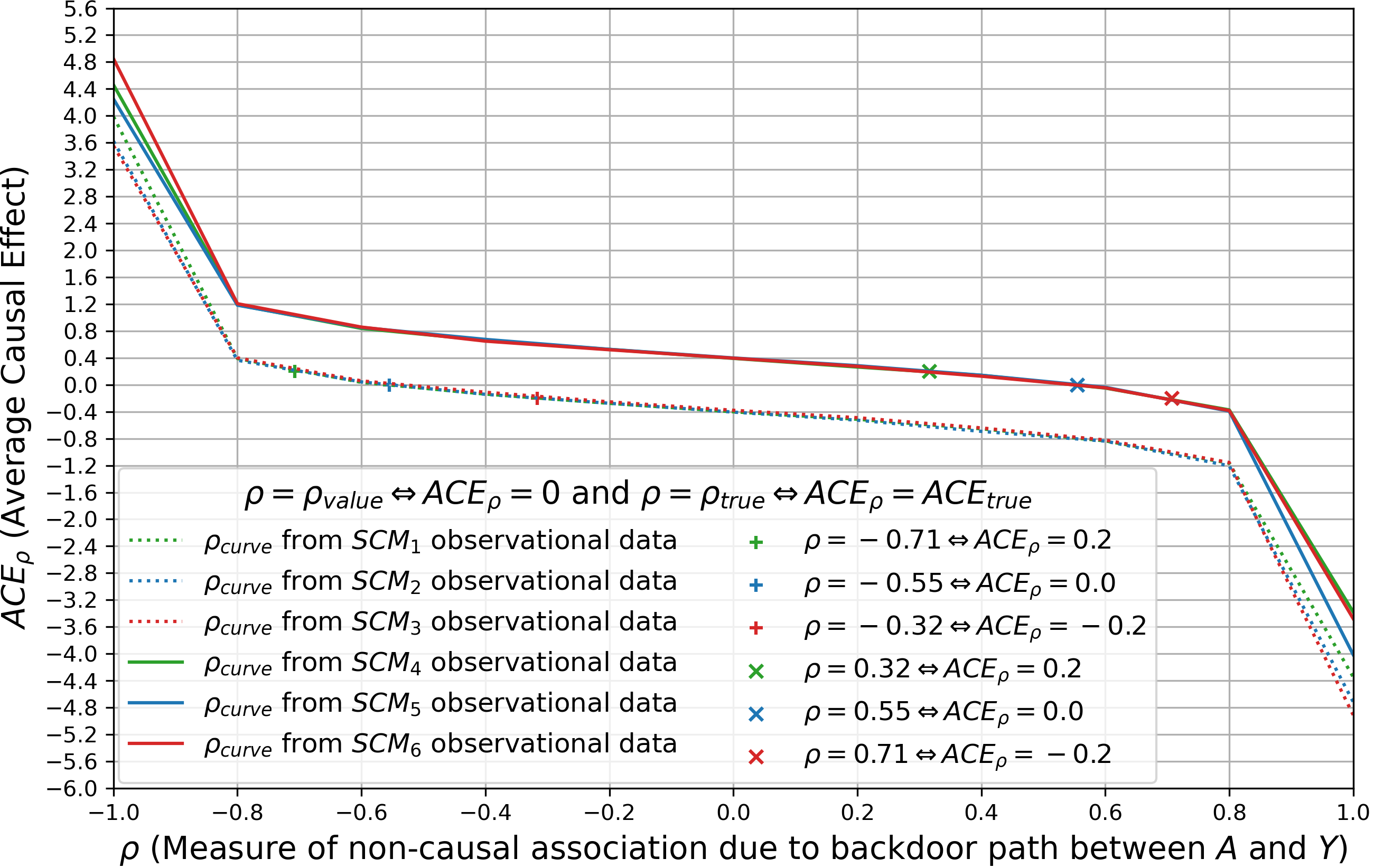}
      \caption{Six $\rho_{curve}$ plots from the observational datasets sampled from the six SCMs in Table \ref{tbl: Jose toy SCM}.}
      \label{sfig:Jose_HVR_SCMs_050000_rho_curves_a}
    \end{subfigure}
    \end{center}
  \caption{Observationally equivalent SCMs, i.e., $\{SCM_{1},SCM_{2},SCM_{3}\}$ and $\{SCM_{4},SCM_{5},SCM_{6}\}$, show similar scatter plots and $\rho_{curve}$ plots.}
  \label{fig:Jose_HVR_SCMs_050000_rho_curves}
\end{figure}

Figure \ref{sfig:Jose_HVR_SCMs_050000_rho_curves_b} shows the scatter plot of each of the six observational datasets obtained by sampling the six SCMs in Table \ref{tbl: Jose toy SCM}. Each dataset contains 50,000 observations. As expected, the SCMs that are observationally equivalent present similar observational data distributions, even though the SCMS are not interventionally equivalent. Likewise, Figure \ref{sfig:Jose_HVR_SCMs_050000_rho_curves_a} shows that the datasets corresponding to observationally equivalent SCMs result in similar $\rho_{curve}$ plots. From this figure, we can note that $\rho{=}\rho_{value}$ implies $ACE_{\rho}{=}0$ and $\rho{=}\rho_{true}$ implies $ACE_{\rho}{=}ACE_{true}$, as can be verified with the help of Table \ref{tbl: Jose toy SCM}. All these observations confirm the accuracy of our $\rho$-GNF for causal inference.

The SCM in Equation \ref{eq:eco} is a simple example involving a continuous outcome and linear relationships, similar to the ones studied in the works \citep{cinelli2019sensitivity, cinelli2020making}. However, our $\rho$-GNF can learn arbitrary non-linear transformations, and thus our sensitivity analysis method also applies to non-linear SCMs, as we demonstrate with further experiments involving binary and categorical outcomes in the next sections. Before that, we show the results of analyzing $SCM_1$ in Table \ref{tbl: Jose toy SCM} with our Bayesian $\rho$-GNF.

\begin{figure}[t!]
  \begin{center}
    \begin{subfigure}[t]{1\textwidth}
      \centering
      \includegraphics[width=1\textwidth]{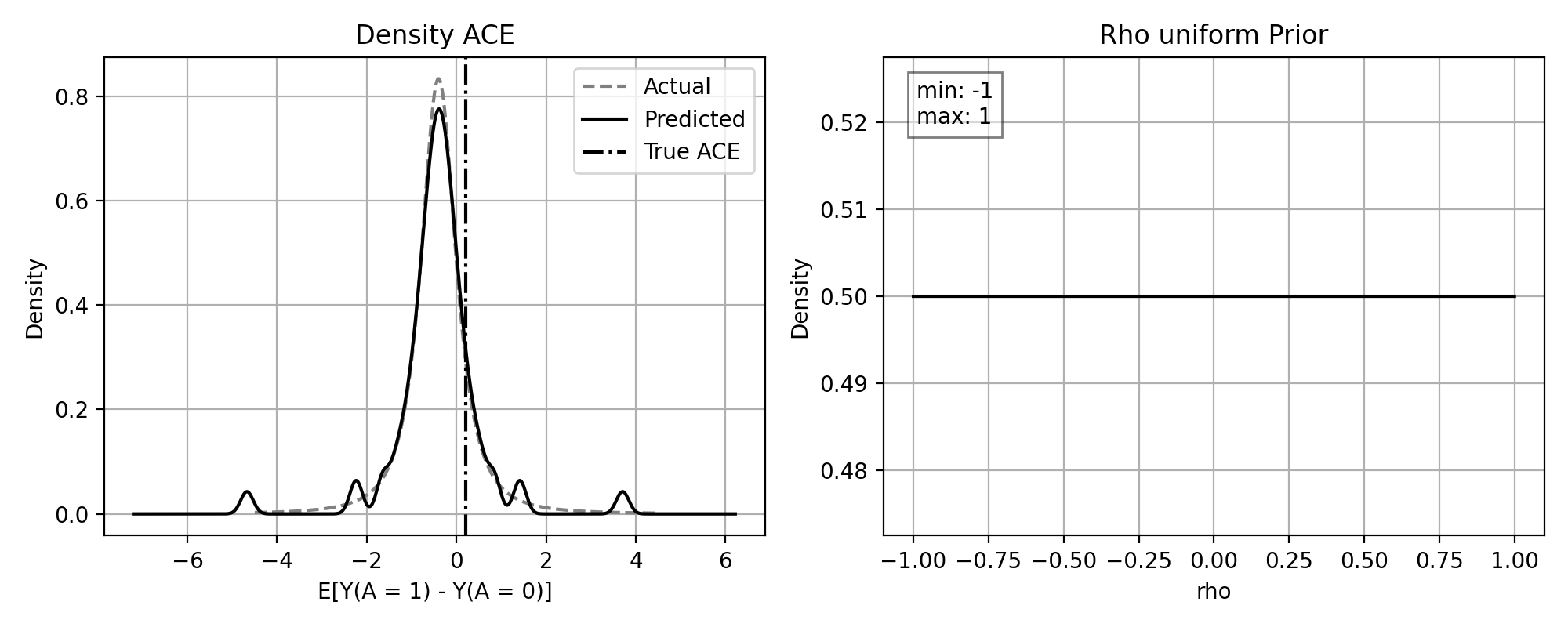}
      \caption{Uniform prior.}
      \label{fig:ace-uniform}
    \end{subfigure}
    \quad
    \begin{subfigure}[t]{1\textwidth}
      \centering
    \vspace{1cm}
      \includegraphics[width=1\textwidth]{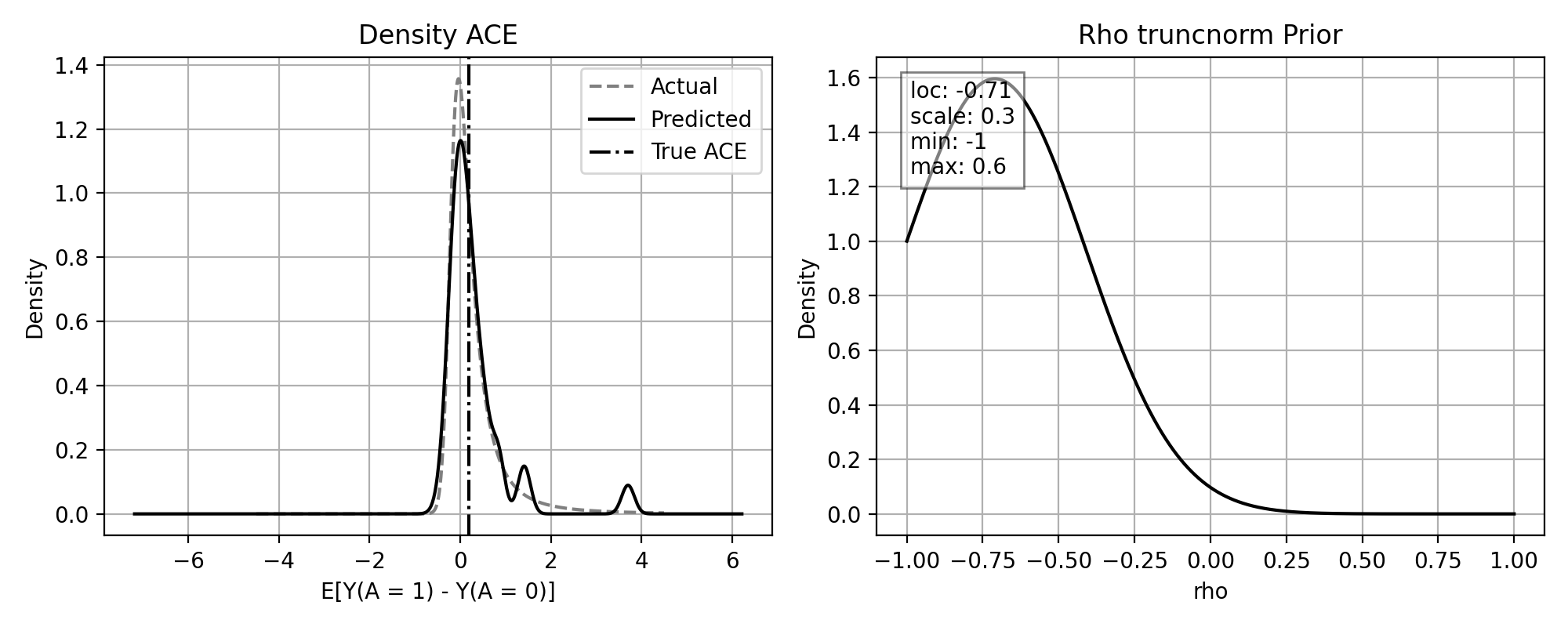}
      \caption{Truncated normal prior.}
      \label{fig:ace-truncnorm}
    \end{subfigure}
   \end{center}
  \caption{Shown on the left is the posterior distribution of the ACE for the $\p$ prior shown on the right. The distribution labeled "actual" is the theoretical convergence limit for $n \rightarrow \infty$ of a universal function approximator like the $\rho$-GNF when trained on $n$ data points assuming the given $\p$ prior. The curve labeled "predicted" shows the distribution estimated with an actual $\rho$-GNF trained on $n = 100,000$ data samples.}
  \label{fig:ace-posterior}
\end{figure}

Specifically, we estimate the posterior distribution of the ACE for two different $\p$ priors. Adopting a uniform distribution over $\p$, Figure \ref{fig:ace-uniform} shows that the ACE appears to be negative. On the other hand, when we adopt a truncated normal distribution that is centered around the true $\p$ value (i.e., $\rho_{true}$ in Table \ref{tbl: Jose toy SCM}), we can see in Figure \ref{fig:ace-truncnorm} that the distribution over the ACE shifts towards the true ACE value (i.e., $\text{ACE}_{true}$ in Table \ref{tbl: Jose toy SCM}).

In practice, researchers might also be interested in estimating the probability for the treatment $A$ having a positive effect on the outcome $Y$. Because our Bayesian $\rho$-GNF provides us with the full posterior distribution of the ACE, we can calculate $P(ACE > 0)$. For the uniform prior we get $P(ACE > 0) = 0.21$ while for the truncated normal prior we get $P(ACE > 0) = 0.61$.

\begin{figure}[t!]
\begin{center}  \includegraphics[width=\linewidth,height=\linewidth,keepaspectratio]{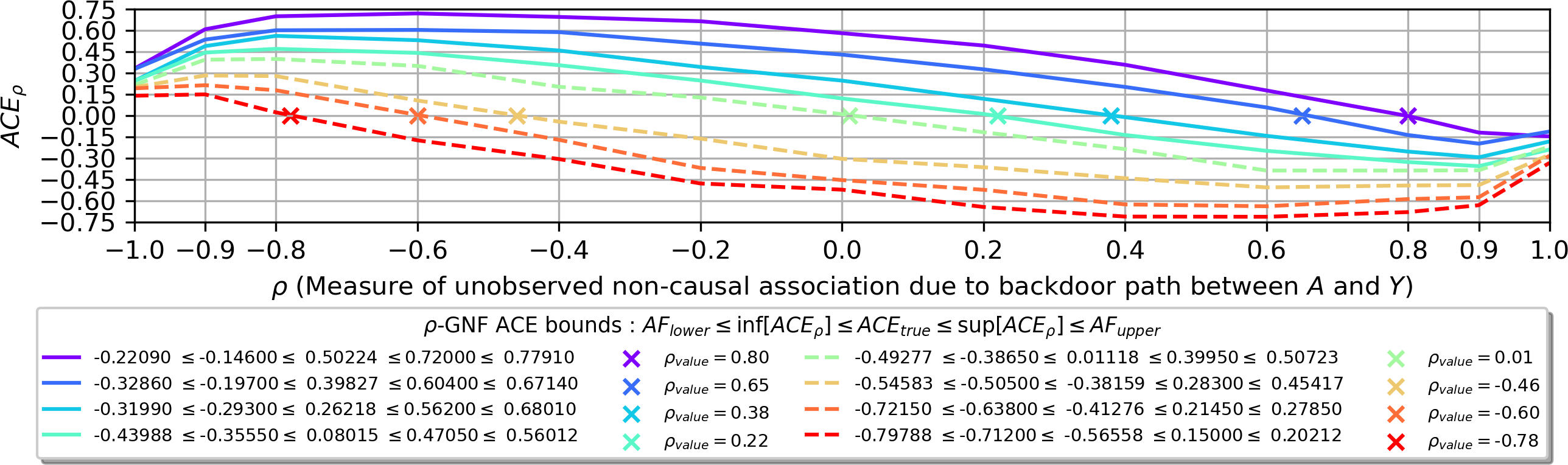}
\end{center}
\caption{Eight $\rho_{curve}$ plots corresponding to eight of the observational datasets in the experiments with binary outcome, and their respective $\rho_{value}$ and bounds.}\label{fig:ndgp_0000_9500_subset}
\end{figure}

\subsection{Experiments with Binary Outcome}\label{sec:experiments simbin}

For our second set of simulated experiments, we consider a setting used in previous works on sensitivity analysis \citep{sjolander2020note, sjolander2021novel, pena2022simple}. Namely, a SCM with binary treatment $A$, binary outcome $Y$, and binary confounder $U$. The SCM is randomly parameterized by sampling $\{P(U), P(A|U), P(Y|A,U)\}$ uniformly from the interval $[0,1]$. We do so 20 times to produce 20 SCMs that are sampled to produce 20 observational datasets over the random vector $(A,Y)$ by discarding the values for $U$. Each dataset contains 50,000 observations. For evaluation purposes, we compute the true ACE (i.e., $ACE_{true}$) for the SCMs sampled by adjusting for $U$ \citep{pearl2009causality}. From the datasets sampled, we can only compute the assumption free (AF) bounds of $ACE_{true}$ as follows, since $U$ is not observed:
\begin{equation}\label{eq:AF}
AF_{lower}{=}q_1p_1{-}q_0p_0{-}p_1 \leq ACE_{true}\leq AF_{upper}{=}q_1p_1{-}q_0p_0{+}p_0
\end{equation}
where $p_a{=}P(A{=}a)$ and $q_a{=}P(Y{=}1|A{=}a)$ are estimated from the data \citep{robins1989analysis, manski1990sensitivitybounds}.

Figure~\ref{fig:ndgp_0000_9500_subset} shows eight $\rho_{curve}$ plots that are representative of the results obtained. Note that for each $\rho_{curve}$, the ACE bounds obtained as the infimum and supremum of the curve (i.e., $\inf[ACE_{\rho}]$ and $\sup[ACE_{\rho}]$) include $ACE_{true}$. As expected due to the Gaussian copula assumption, these bounds are narrower than the AF bounds. These observations together with the ones made in the previous section confirm that our $\rho$-GNF is accurate for both continuous and binary outcomes. This generality distinguishes our method from the existing sensitivity analysis methods. The next section presents experiments with a categorical outcome and real-world data.

\subsection{Experiments with Real-World Datasets}\label{sec:real}

In this section, we present experiments with two real-world datasets. First, we use our $\rho$-GNF approach to analyze the impact of the International Monetary Fund on child poverty \citep{DAOUD2024102973,Daoud2017IMFchildhealth,hallerod2013badgovernance_poorchildren,balgi2022counterfactual}. Afterwards, we present a sensitivity analysis for potential outcomes with the classical Blau and Duncan social mobility dataset \citep{blau1967american}.

\begin{figure}[t!]
\begin{center} \includegraphics[width=\linewidth,height=\linewidth,keepaspectratio]{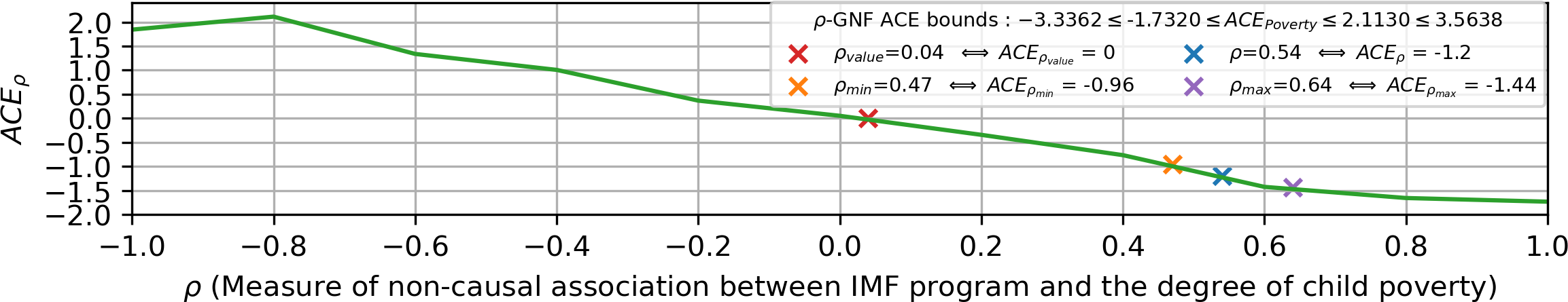}
\end{center}
\caption{$\rho_{curve}$ for the degree of the child poverty, and its $\rho_{value}$ and bounds.}\label{fig:IMFCP_sensitivity_analysis_full}
\end{figure}

\subsubsection{International Monetary Fund}\label{sec:experimentsIMF}

The International Monetary Fund (IMF) is an international organization that aims to promote global macroeconomic stability through a series of programs. However, the impact of these programs on children is a subject of debate \citep{Daoud2017IMFchildhealth, daoud2018structuraladjustments, DAOUD2019foodagri}. In this section, we study the impact of the programs (treatment $A$) on child poverty (outcome $Y$). We consider the IMF child poverty dataset used previously in the works \citep{DAOUD2024102973,Daoud2017IMFchildhealth,hallerod2013badgovernance_poorchildren,balgi2022counterfactual}. It contains 1,941,734 observations each corresponding to a child under the age of 18 residing in 67 countries from the Global-South region, which includes the least developed countries.\footnote{Due to the sensitive nature and the accompanying ethical considerations, the dataset is not publicly available but it can be requested upon from the original authors.} For each child, the dataset records whether she receives the treatment (i.e., she lives in a country adopting the IMF program) as well as her degree of poverty. The degree of poverty ranges from 0 (no poverty) to 7 (severe poverty). It is calculated as the sum of seven binary indicators of poverty, representing access to education, health services, information, sanitation, shelter, food and water.

It is most likely that the IMF dataset is subject to unmeasured confounding, and thus the true ACE is not identifiable from the data \cite{pearl2009causality}. However, we can bound it by computing the AF bounds in Equation \ref{eq:AF} for each of the seven binary indicators of poverty, and then summing them. This results in $AF_{lower}=-3.34$ and $AF_{upper}=3.56$.

Figure \ref{fig:IMFCP_sensitivity_analysis_full} shows the $\rho_{curve}$ produced by our sensitivity analysis method. As expected due to the Gaussian copula assumption, the bounds $\inf[ACE_{\rho}]=-1.73$ and $\sup[ACE_{\rho}]=2.11$ obtained from the curve are narrower than the AF bounds. The fact that the $\rho_{value}$ is as small as 0.04 indicates that the unmeasured confounding and the causal effect cancel each other, i.e., they have equal magnitude but opposite signs. In other words, either there is no unmeasured confounding and no causal effect of the IMF program on child poverty, or both exist and have the same strength but opposite signs. Previous works rule out the first explanation \citep{DAOUD2024102973,Daoud2017IMFchildhealth,hallerod2013badgovernance_poorchildren,balgi2022counterfactual}. Moreover, the second explanation is not unreasonable as we elaborate next. A country with severe socioeconomic turmoil (i.e., large $\varepsilon_Y$) causing severe child poverty is associated with higher incentives to apply for the IMF program (i.e., large $\varepsilon_A$) to overcome the socioeconomic turmoil. Similarly, a country with less socioeconomic turmoil (i.e., small $\varepsilon_Y$) has less incentives to apply for the IMF program (i.e., small $\varepsilon_A$). Therefore, it is not unreasonable to assume that $\varepsilon_A$ and $\varepsilon_Y$ are positively associated, which correspond to assuming a positive $\rho$ value in our sensitivity analysis framework. In that case, the true ACE would be negative as can be seen in Figure \ref{fig:IMFCP_sensitivity_analysis_full}. For instance, if the domain experts were to believe that $\rho \in [0.47, 0.64]$, then $ACE_{true} \in [-1.44,-0.96]$ which indicates that the IMF program is beneficial for reducing child poverty. This real-world example illustrates how our $\rho$-GNF helps the analyst to obtain ACE bounds that are more informative than the AF bounds. Our Bayesian $\rho$-GNF can provide the analyst with a even more detailed picture, as we show below.

\begin{figure}[t!]
\begin{center} \includegraphics[width=\linewidth]{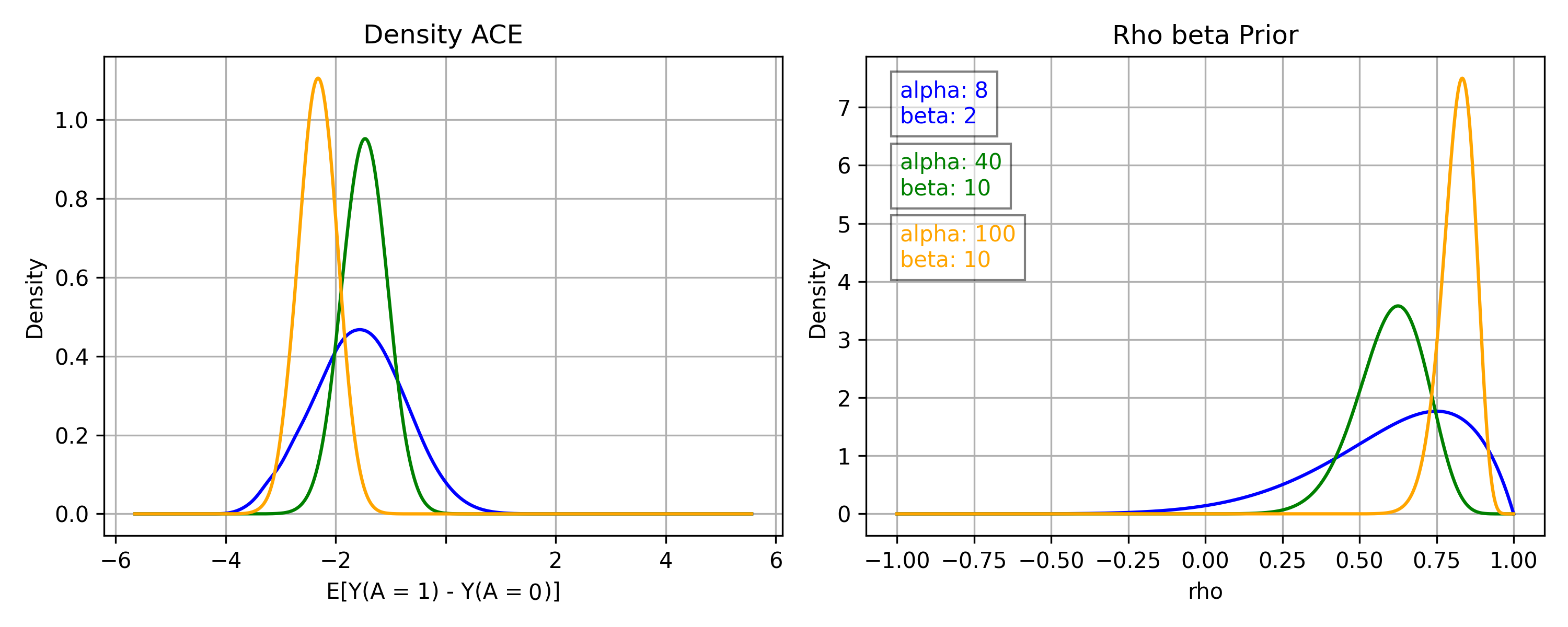}
\end{center}
\caption{Estimated posterior distributions of the ACE assuming a scaled and shifted beta distribution as $\p$ prior with different values for $\alpha$ and $\beta$.}\label{fig:IMF_beta_prior}
\end{figure}

Specifically, since $\varepsilon_A$ and $\varepsilon_Y$ are arguably positively associated, we choose a scaled and shifted beta distribution with three different parameter combinations of $\alpha$ and $\beta$ that all have positive mean as $\p$ prior. Figure \ref{fig:IMF_beta_prior} shows the prior distributions and the resulting posterior distributions of the ACE. Comparing the blue and green posterior distributions that have the same mean but different variance, we can conclude from the plot that the more certain one is in the value of $\p$, the smaller the variance of the ACE. Comparing the green to the yellow posterior distributions, we can also conclude that shifting the mean of the prior towards larger values of $\p$ implies lower ACE values. Our Bayesian analysis affirms the beneficial nature of the IMF program in the reduction of child poverty, as for all three prior distributions with positive mean, most of the posterior density lies on negative ACE values. For example, the blue prior with $\alpha = 8$ and $\beta = 2$ results in a 95\% credible interval for the ACE of $[-3.177, -0.057]$.

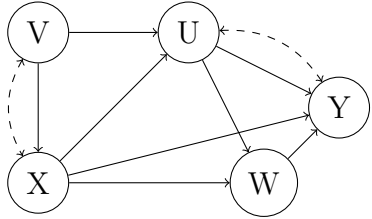
\begin{figure}[t!]
    \centering
    \begin{tikzpicture}
        \node[draw, circle] (X) at (0,0) {X};
        \node[draw, circle] (W) at (3,0) {W};
        \node[draw, circle] (Y) at (4,1) {Y};
        \node[draw, circle] (U) at (2,2) {U};
        \node[draw, circle] (V) at (0,2) {V};
        
        \draw[->] (X) -- (W);
        \draw[->] (X) -- (Y);
        \draw[->] (X) -- (U);
        \draw[->] (W) -- (Y);
        \draw[->] (U) -- (Y);
        \draw[->] (U) -- (W);
        \draw[<->, dashed, bend left] (U) to (Y);
        \draw[<->, dashed, bend right] (V) to (X);
        \draw[->] (V) -- (U);
        \draw[->] (V) -- (X);
    \end{tikzpicture}
    \caption{The causal graph corresponding to the Blau and Duncan dataset, except for the bidirected edge between $U$ and $Y$. Bidirectional edges indicate correlation between the corresponding unmodeled $\varepsilon$ nodes. The nodes in the graph represent the father’s educational attainment ($V$), father’s occupational status ($X$), son’s educational attainment ($U$), the occupational status of the son’s first job ($W$), and the occupational status of the son’s job in 1962 ($Y$), i.e., the year the data were collected.}
    \label{fig:blau-duncan-dag}
\end{figure}

\subsubsection{Blau and Duncan Social Mobility Dataset}

The work \citep{blau1967american} is considered a pioneering study of intergenerational social mobility in the U.S. The authors used linear path analysis (i.e., linear SCMs) to analyze data from the 1962 “Occupational Changes in a Generation” survey of 20,000 men aged 20–64. They examined how family background, education, and early career achievements influenced occupational outcomes. The study assumed the causal graph depicted in Figure \ref{fig:blau-duncan-dag}, except for the bidirected edge between $U$ and $Y$.

The focus of the original study was on $Y$, i.e., the occupational status of the son’s job in 1962 when the survey was conducted. The original causal graph did not contain any hidden confounding involving $Y$. However, as argued in the work \citep{Balgi2025}, it is not unreasonable to assume that the son's educational attainment $U$ and $Y$ are confounded by motivation or grit, which is not measured in the dataset.\footnote{Note that the hidden confounder is typically denoted as $U$ in the literature, while we stick to the notation of the original paper where $U$ is the son's educational attainment.} This is represented by adding the bidirected edge between $U$ and $Y$ to the original causal graph in Figure \ref{fig:blau-duncan-dag}. Both $U$ and $Y$ are categorical random variables. The former has categories 0-8, and the latter 0-96.

\begin{figure}[t!]
\begin{center}
    \includegraphics[width=0.9\textwidth]{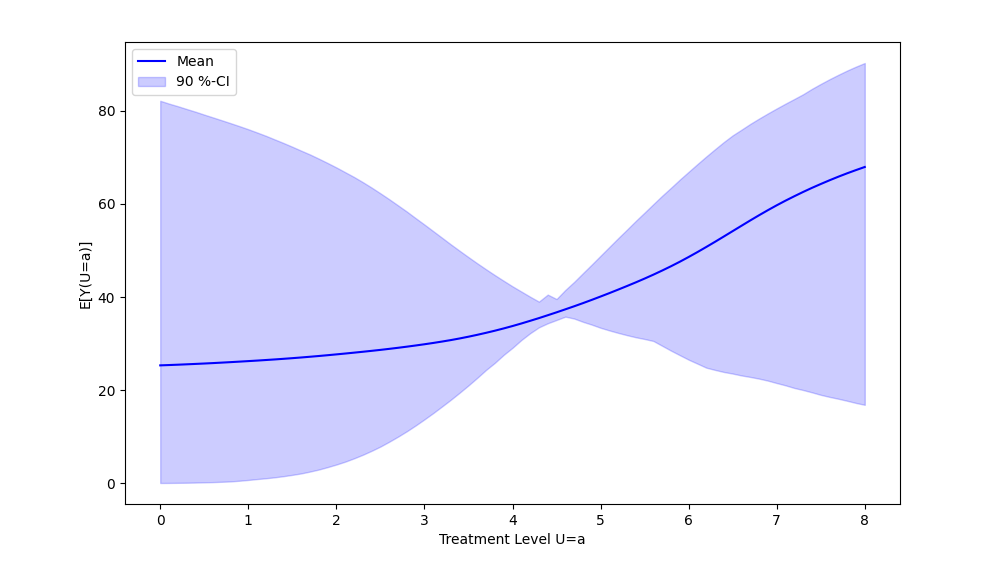}
\end{center}
\caption{Posterior mean of the potential outcome of $Y$ and $90\ \%$ credible interval assuming a uniform $\p$ prior between $-1$ and $1$.}\label{fig:blau-duncan-potential-outcome-uniform}
\end{figure}

We run our Bayesian $\rho$-GNF framework to estimate the expected potential outcome of $Y$ for different treatment levels of $U$, as a function of the strength of their unmeasured confounder. As discussed above, this confounder may well be motivation or grit. However, it may be motivation just for educational achievements, or just for work achievements, or both, or none. In other words, we are uncertain about the sign of the influence of this confounder on $U$ and $Y$, and thus we adopt a uniform $\rho$ prior in the interval $[-1,1]$ following the recommendations in Section \ref{sec:prior}. We compute the posterior distribution of the potential outcome of $Y$ for different treatment levels of $U$. Figure \ref{fig:blau-duncan-potential-outcome-uniform} plots the posterior mean and $90 \%$ credible intervals. We can conclude that $U$ seems to have an effect on $Y$, and this effect seems to be non-linear. Moreover, the effect seems to be rather insensitive towards confounding at treatment levels around 4-5 but quite sensitive at higher and lower treatment levels. This is partially explained by the fact that the more extreme the treatment level is the fewer the observations in the dataset. This example illustrates that our Bayesian $\rho$-GNF framework is an informative tool for sensitivity analysis with categorical treatment and/or outcomes. This together with our previous experiments with continuous and binary outcomes demonstrate the generality of our framework, which is a feature that distinguishes it from the existing sensitivity analysis methods.

\section{Conclusion}\label{sec:conclusion}

We proposed $\rho$-GNF, a novel approach for sensitivity analysis to unobserved confounding that is based on copula-based normalizing flows. Our approach contains a bounded and interpretable sensitivity parameter $\rho$ representing the unobserved non-causal association between the observed treatment and outcome due to confounding. Under the Gaussian copula assumption, we showed that $\rho$-GNF enabled us to estimate the causal effect as a function of $\rho$ in the form of the $\rho_{curve}$. The $\rho_{curve}$ enabled us to identify the $\rho_{value}$, i.e., the confounding strength needed to nullify the causal effect. This is related to the E-value in the literature. The $\rho_{curve}$ also enabled us to identify empirical bounds of the causal effect that are narrower than the assumption free bounds in the literature.

We also proposed a Bayesian version of $\rho$-GNF by defining a prior distribution over the sensitivity parameter $\p$, which allowed us to calculate the posterior distribution over any quantity of interest that can be deducted from the $\rho$-GNF. This enabled us to derive credible intervals for the causal effect. We also discussed how to choose a suitable prior based on expert knowledge about the hidden confounder.

Finally, we illustrated the benefits of our sensitivity analysis method with simulated and real-world data under different settings, e.g., with continuous, binary and categorical outcomes. This generality distinguishes our method from the methods in the literature.

It is worth recalling from the discussion in Section \ref{sec:assumption} that, although the Gaussian copula fits our framework particularly well, any other copula may be used. As a matter of fact, we recommend using several copulas and integrate the conclusions drawn from all of them, in order to increase robustness. Likewise, we recommend integrating too the results of other sensitivity analysis methods in the literature, which are based on different assumptions than ours. This process is sometimes called triangulation. We do not pursue this further in this paper because it is a known practice \citep{hernan2009ipw}.

It is also worth commenting on the scalability of the framework presented in this paper to large datasets. Our experiments in Section \ref{sec:experimentsIMF} with a dataset of almost 2 million observations were run in an ordinary computer, since no particularly deep neural networks were required. Specifically, the neural networks used had three fully connected hidden layers of between 5 and 20 units each. This proves that our framework scales well.

Finally, it is worth reminding the reader of the following caveat against over-interpreting causal conclusions, including the ones in this work. Sensitivity analysis rarely rules out the possibility that the causal effect is null, i.e., that the observed association between exposure and outcome is not solely due to confounding. Instead, sensitivity analysis informs the user of the confounding strength required to nullify the causal effect, and it is up to the user to decide whether such strength is likely or not in the domain at hand. In our Bayesian framework, we similarly let the user specify a prior distribution over the confounding strength and produce a posterior distribution over the causal effect to decide if the null causal effect is likely or not. Therefore, any causal claim is based on the assumptions made by the sensitivity analysis method considered and the final judgment of the user about the confounding strength required to nullify the causal effect. As mentioned above, one way to increase the strength of causal claims is by triangulation \citep{hernan2009ipw}, i.e., by integrating the results of several sensitivity analysis methods based on potentially different assumptions.

\appendix
\section{Proof of Proposition \ref{prop:equivalent-scms}}
\label{proposition}

First, note that $\delta$ needs to be bound between $\pm \sqrt{\frac{(1 - \p^2) \gamma^2}{\p^2}}$ as otherwise $\frac{1}{1 - \p^2} - \frac{\gamma^2 + \delta^2}{\gamma^2} \cdot \frac{\p^2}{1 - \p^2}$ would be negative, and consequently $\tau$ would not be a real number. Likewise, $\gamma$ needs to be non-zero as otherwise $\delta$ or $\Tilde{\varepsilon}_A$ are $0/0$. Furthermore, observe that when
\begin{equation*}
        \vectwod{\frac{1}{\sqrt{\gamma^2 + \delta^2}} \left( \gamma U + \delta \Tilde{\varepsilon}_A \right)}{\lambda \p U + \tau \sqrt{1-\p^2} \Tilde{\varepsilon}_Y} = \vectwod{Z_A}{Z_Y}
        \sim
        \Normal{\vectwod{0}{0}}{\begin{pmatrix}
            1 & \p \\
            \p & 1
        \end{pmatrix}}
\end{equation*}
then the SCMs in Equation \ref{seq:eps dist scm1} and in the proposition are observationally and interventionally equivalent. We know that because $\Tilde{\varepsilon}_A, \Tilde{\varepsilon}_Y, \varepsilon_U \simiid \Normal{0}{1}$, any linear combination of them is also normally distributed. It therefore suffices to show that the mean vector is $(0, 0)^T$, the variances are $1$, and the covariance is $\p$.

First, note that
    \begin{align*}
        \E{\frac{1}{\sqrt{\gamma^2 + \delta^2}} \left( \gamma U + \delta \Tilde{\varepsilon}_A \right)}
        &= \frac{1}{\sqrt{\gamma^2 + \delta^2}} \left( \gamma \E{U} + \delta \E{\Tilde{\varepsilon}_A } \right) \\
        &= \frac{1}{\sqrt{\gamma^2 + \delta^2}} \left( \gamma \cdot 0 + \delta \cdot 0 \right) \\
        &= 0 \enspace .
    \end{align*}
    
Second, note that 
    \begin{align*}
        \E{\lambda \p U +\tau \sqrt{1-\p^2} \Tilde{\varepsilon}_Y}
        &= \lambda \p \E{U} +\tau \sqrt{1-\p^2} \E{\Tilde{\varepsilon}_Y} \\
        &= \lambda \p \cdot 0 +\tau \sqrt{1-\p^2} \cdot 0 \\
        &= 0 \enspace .
    \end{align*}
    
Third, note that
    \begin{align*}
        \var{\frac{1}{\sqrt{\gamma^2 + \delta^2}} \left( \gamma U + \delta \Tilde{\varepsilon}_A \right)}
        &= \frac{1}{\gamma^2 + \delta^2} \left( \gamma^2 \var{U} + \delta^2 \var{\Tilde{\varepsilon}_A } \right) \\
        &= \frac{\gamma^2 + \delta^2}{\gamma^2 + \delta^2} \\
        &= 1 \enspace .
    \end{align*}
    
Fourth, note that   
    \begin{align*}
        \var{\lambda \p U +\tau \sqrt{1-\p^2} \Tilde{\varepsilon}_Y}
        &= \lambda^2 \p^2 \var{U} +\tau^2 (1 - \p^2) \var{\Tilde{\varepsilon}_Y} \\
        &= \lambda^2 \p^2 +\tau^2 (1 - \p^2) \\
        &= \frac{\gamma^2 + \delta^2}{\gamma^2} \p^2 + \left( \frac{1}{1 - \p^2} - \frac{\gamma^2 + \delta^2}{\gamma^2} \cdot \frac{\p^2}{1 - \p^2} \right) \cdot (1 - \p^2) \\
        &= \frac{(\gamma^2 + \delta^2) \p^2}{\gamma^2} + \left( 1 - \frac{(\gamma^2 + \delta^2) \p^2}{\gamma^2} \right)\\
        &= 1 \enspace .
    \end{align*}
    
Finally, note that  
    \begin{align*}
        \covar{\frac{1}{\sqrt{\gamma^2 + \delta^2}} \left( \gamma U + \delta \Tilde{\varepsilon}_A \right)}{\lambda \p U +\tau \sqrt{1-\p^2} \Tilde{\varepsilon}_Y}
        &= \frac{\gamma \cdot \p \cdot \lambda}{\sqrt{\gamma^2 + \delta^2}} \covar{U}{U}\\
        &+  \frac{\gamma \cdot \sqrt{1 - \p^2} \cdot\tau}{\sqrt{\gamma^2 + \delta^2}} \covar{U}{\Tilde{\varepsilon}_Y} \\
        &+ \frac{\lambda \cdot \delta \cdot \p}{\sqrt{\gamma^2 + \delta^2}} \covar{\Tilde{\varepsilon}_A}{U} \\
        &+  \frac{\delta \cdot \tau \cdot \sqrt{1 - \p^2}}{\sqrt{\gamma^2 + \delta^2}} \covar{\Tilde{\varepsilon}_A}{\Tilde{\varepsilon}_Y} \\
        &= \frac{\gamma}{\sqrt{\gamma^2 + \delta^2}} \cdot \lambda \cdot \p \\
        &= \frac{\gamma}{\sqrt{\gamma^2 + \delta^2}} \cdot \frac{\sqrt{\gamma^2 + \delta^2}}{\gamma} \cdot \p \\
        &= \p \enspace .
    \end{align*}

\section{Proof of Theorem \ref{thm:sign-of-rho}}
\label{theorem}

Recall that $T_A^{-1}(Z_A)$ is strictly increasing in $Z_A$ by construction. Then, we have that
    \begin{align*}
        \sgn{\partialDer{\Tilde{t}_A}{U}} &= \sgn{\partialDer{Z_A}{U} \cdot \derivative{T_A^{-1}(Z_A)}{Z_A}} \\
        &= \sgn{\frac{\gamma}{\sqrt{\gamma^2 + \delta^2}}} \cdot \sgn{\derivative{T_A^{-1}(Z_A)}{Z_A}} \\
        &= \sgn{\frac{\gamma}{\sqrt{\gamma^2 + \delta^2}}} \cdot 1 \\
        &= \sgn{\gamma} \enspace .
    \end{align*}

Recall that $T_{Y|A}^{-1}(Z_Y)$ is strictly increasing in $Z_Y$ by construction. Then, we have that
    \begin{align*}
        \sgn{\partialDer{\Tilde{t}_Y}{U}} &= \sgn{\partialDer{Z_Y}{U} \cdot \derivative{T_{Y|A}^{-1}(Z_Y)}{Z_Y}} \\
        &= \sgn{\lambda \p} \cdot \sgn{\derivative{T_{Y|A}^{-1}(Z_Y)}{Z_Y}} \\
        &= \sgn{\p \frac{\sqrt{\gamma^2 + \delta^2}}{\gamma}} \cdot 1 \\
        &= \sgn{\p} \cdot \sgn{\gamma} \enspace .
    \end{align*}

Since $\gamma\neq 0$ by definition, we then have that
\begin{align*}
        \p > 0 &\implies \sgn{\partialDer{\Tilde{t}_A}{U}} = \sgn{\partialDer{\Tilde{t}_Y}{U}} \\
        \p \leq 0 &\implies \sgn{\partialDer{\Tilde{t}_A}{U}} \neq \sgn{\partialDer{\Tilde{t}_Y}{U}} \enspace .
    \end{align*}

\bibliographystyle{elsarticle-num} 
\bibliography{pgm2024_short}

\begin{thebibliography}{10}
\expandafter\ifx\csname url\endcsname\relax
  \def\url#1{\texttt{#1}}\fi
\expandafter\ifx\csname urlprefix\endcsname\relax\def\urlprefix{URL }\fi
\expandafter\ifx\csname href\endcsname\relax
  \def\href#1#2{#2} \def\path#1{#1}\fi

\bibitem{wright1921correlationcausation}
S.~Wright, \href{https://naldc.nal.usda.gov/download/IND43966364/pdf}{Correlation and Causation}, Journal of Agricultural Research 20 (1921) 557--585.

\bibitem{fisher1936designofexperiments}
R.~A. Fisher, \href{https://doi.org/10.1136/bmj.1.3923.554-a}{Design of Experiments}, British Medical Journal 1 (1936) 554--554.

\bibitem{cox1958planning}
D.~R. Cox, \href{https://doi.org/10.1017/S0020268100038063}{Planning of Experiments}, Wiley, 1958.

\bibitem{Imbens2015CIinSocialscience}
G.~W. Imbens, D.~B. Rubin, \href{https://doi.org/10.1017/cbo9781139025751}{Causal Inference for Statistics, Social, and Biomedical Sciences: An Introduction}, Cambridge University Press, 2015.

\bibitem{pearl2009causality}
J.~Pearl, \href{https://dl.acm.org/doi/book/10.5555/1642718}{Causality: Models, Reasoning and Inference}, Cambridge University Press, 2009.

\bibitem{hernan2009ipw}
M.~A. Hern{\'a}n, J.~M. Robins, \href{https://cdn1.sph.harvard.edu/wp-content/uploads/sites/1268/2022/10/hernanrobins_WhatIf_15sep22.pdf}{Causal Inference: What If}, Chapman \& Hall/CRC, 2020.

\bibitem{schlesselman1978assessing}
J.~J. Schlesselman, \href{https://doi.org/10.1093/oxfordjournals.aje.a112581}{Assessing Effects of Confounding Variables}, American Journal of Epidemiology 108~(1) (1978) 3--8.

\bibitem{manski1990sensitivitybounds}
C.~F. Manski, \href{http://www.jstor.org/stable/2006592}{Nonparametric Bounds on Treatment Effects}, American Economic Review 80~(2) (1990) 319--323.

\bibitem{imbens2003sensitivity}
G.~W. Imbens, \href{https://doi.org/10.1257/000282803321946921}{Sensitivity to Exogeneity Assumptions in Program Evaluation}, American Economic Review 93~(2) (2003) 126--132.

\bibitem{brumback2004sensitivity}
B.~A. Brumback, M.~A. Hern{\'a}n, S.~J. Haneuse, J.~M. Robins, \href{https://doi.org/10.1002/sim.1657}{Sensitivity Analyses for Unmeasured Confounding Assuming a Marginal Structural Model for Repeated Measures}, Statistics in Medicine 23~(5) (2004) 749--767.

\bibitem{vanderweele2011bias}
T.~J. VanderWeele, O.~A. Arah, \href{https://doi.org/10.1097/ede.0b013e3181f74493}{Bias Formulas for Sensitivity Analysis of Unmeasured Confounding for General Outcomes, Treatments, and Confounders}, Epidemiology (2011) 42--52.

\bibitem{cornfield1959smoking}
J.~Cornfield, W.~Haenszel, E.~C. Hammond, A.~M. Lilienfeld, M.~B. Shimkin, E.~L. Wynder, \href{https://doi.org/10.1093/ije/dyp289}{Smoking and Lung Cancer: Recent Evidence and a Discussion of Some Questions}, Journal of the National Cancer Institute 22~(1) (1959) 173--203.

\bibitem{cochran1973controlling}
W.~G. Cochran, D.~B. Rubin, \href{https://www.jstor.org/stable/25049893}{Controlling Bias in Observational Studies: A Review}, Sankhy{\=a}: The Indian Journal of Statistics, Series A (1973) 417--446.

\bibitem{rothman2008modern}
K.~J. Rothman, S.~Greenland, T.~L. Lash, Modern Epidemiology, Wolters Kluwer Health/Lippincott Williams \& Wilkins, 2008.

\bibitem{lash2009applying}
T.~L. Lash, M.~P. Fox, A.~K. Fink, et~al., \href{https://doi.org/10.1007/978-0-387-87959-8}{Applying Quantitative Bias Analysis to Epidemiologic Data}, Springer, 2009.

\bibitem{wehenkel2020GNF}
A.~Wehenkel, G.~Louppe, \href{http://proceedings.mlr.press/v130/wehenkel21a/wehenkel21a.pdf}{Graphical Normalizing Flows}, in: International Conference on Artificial Intelligence and Statistics (AISTATS), 2021, pp. 37--45.

\bibitem{balgi2022cgnf}
S.~Balgi, J.~M. Pe{\~n}a, A.~Daoud, \href{https://ojs.aaai.org/index.php/AAAI/article/view/21437}{Personalized Public Policy Analysis in Social Sciences Using Causal-Graphical Normalizing Flows}, in: AAAI Conference on Artificial Intelligence (AAAI), 2022, pp. 11810--11818.

\bibitem{JavaloySV23}
A.~Javaloy, P.~S{\'{a}}nchez{-}Mart{\'{\i}}n, I.~Valera, \href{https://doi.org/10.48550/arXiv.2306.05415}{Causal Normalizing Flows: From Theory to Practice}, in: Neural Information Processing Systems (NeurIPS), 2023, pp. 58833--58864.

\bibitem{nelsen2007introduction}
R.~B. Nelsen, \href{https://doi.org/10.1007/0-387-28678-0}{An Introduction to Copulas}, Springer, 2007.

\bibitem{vanderweele2017sensitivity}
T.~J. VanderWeele, P.~Ding, \href{https://doi.org/10.7326/M16-2607}{Sensitivity Analysis in Observational Research: Introducing the {E}-value}, Internal Medicine 167~(4) (2017) 268--274.

\bibitem{robins1989analysis}
J.~M. Robins, \href{https://cdn1.sph.harvard.edu/wp-content/uploads/sites/343/2013/03/nchsr.pdf}{The Analysis of Randomized and Non-randomized AIDS Treatment Trials Using a New Approach to Causal Inference in Longitudinal Studies}, Health Service Research Methodology: A Focus on AIDS (1989) 113--159.

\bibitem{sjolander2020note}
A.~Sj{\"o}lander, \href{https://doi.org/10.1515/jci-2020-0012}{A Note on a Sensitivity Analysis for Unmeasured Confounding, and the Related {E}-value}, Journal of Causal Inference 8~(1) (2020) 229--248.

\bibitem{sjolander2021novel}
A.~Sj{\"o}lander, O.~H{\"o}ssjer, \href{https://doi.org/10.1515/jci-2021-0024}{Novel Bounds for Causal Effects Based on Sensitivity Parameters on the Risk Difference Scale}, Journal of Causal Inference 9~(1) (2021) 190--210.

\bibitem{pena2022simple}
J.~M. Pe{\~n}a, \href{https://doi.org/10.1515/jci-2021-0041}{Simple Yet Sharp Sensitivity Analysis for Unmeasured Confounding}, Journal of Causal Inference 10~(1) (2022) 1--17.

\bibitem{cinelli2019sensitivity}
C.~Cinelli, D.~Kumor, B.~Chen, J.~Pearl, E.~Bareinboim, \href{http://proceedings.mlr.press/v97/cinelli19a/cinelli19a.pdf}{Sensitivity Analysis of Linear Structural Causal Models}, in: International Conference on Machine Learning (ICML), 2019, pp. 1252--1261.

\bibitem{cinelli2020making}
C.~Cinelli, C.~Hazlett, \href{http://dx.doi.org/10.1111/rssb.12348}{ Making Sense of Sensitivity: Extending Omitted Variable Bias}, Journal of the Royal Statistical Society: Series B 82~(1) (2020) 39--67.

\bibitem{balgi2022rho}
S.~Balgi, J.~M. Pe{\~n}a, A.~Daoud, {$\rho$-GNF : A Copula-based Sensitivity Analysis to Unobserved Confounding Using Normalizing Flows}, in: International Conference on Probabilistic Graphical Models (PGM), 2024, pp. 20--37.

\bibitem{ding2016sensitivity}
P.~Ding, T.~J. VanderWeele, \href{https://doi.org/10.1097/ede.0000000000000457}{Sensitivity Analysis Without Assumptions}, Epidemiology 27~(3) (2016) 368.

\bibitem{veitch2020sense}
V.~Veitch, A.~Zaveri, \href{https://proceedings.neurips.cc/paper/2020/hash/7d265aa7147bd3913fb84c7963a209d1-Abstract.html}{Sense and Sensitivity Analysis: Simple Post-hoc Analysis of Bias Due to Unobserved Confounding}, in: Neural Information Processing Systems (NeurIPS), 2020, pp. 10999--11009.

\bibitem{sjolander2022values}
A.~Sj{\"o}lander, S.~Greenland, {\href{https://doi.org/10.1093/ije/dyac018}{Are {E}-values Too Optimistic or Too Pessimistic? Both and Neither!}}, International Journal of Epidemiology (2022).

\bibitem{ilse2021efficient}
M.~Ilse, P.~Forr{\'e}, M.~Welling, J.~M. Mooij, \href{https://doi.org/10.48550/arXiv.2103.04786}{Combining Interventional and Observational Data Using Causal Reductions}, arXiv preprint arXiv:2103.04786 (2021).

\bibitem{ioannidis2019limitations}
J.~Ioannidis, Y.~Tan, M.~Blum, \href{https://doi.org/10.7326/M18-2159}{Limitations and Misinterpretations of {E}-values for Sensitivity Analyses of Observational Studies}, Internal Medicine 170~(2) (2019) 108--111.

\bibitem{tchetgen2012semiparametric}
E.~J.~T. Tchetgen, I.~Shpitser, \href{https://doi.org/10.1214/12-AOS990}{Semiparametric Theory for Causal Mediation Analysis: Efficiency Bounds, Multiple Robustness, and Sensitivity Analysis}, Annals of Statistics 40~(3) (2012) 1816.

\bibitem{lindmark2018sensitivity}
A.~Lindmark, X.~de~Luna, M.~Eriksson, \href{https://doi.org/10.1002/sim.7620}{Sensitivity Analysis for Unobserved Confounding of Direct and Indirect Effects Using Uncertainty Intervals}, Statistics in Medicine 37~(10) (2018) 1744--1762.

\bibitem{peters2017eci}
J.~Peters, D.~Janzing, B.~Sch{\"o}lkopf, \href{https://library.oapen.org/bitstream/handle/20.500.12657/26040/11283.pdf}{Elements of Causal Inference: Foundations and Learning Algorithms}, The MIT Press, 2017.

\bibitem{angus1994probability}
J.~E. Angus, \href{https://doi.org/10.1137/1036146}{The Probability Integral Transform and Related Results}, SIAM Review 36~(4) (1994) 652--654.

\bibitem{mooij2016distinguishing}
J.~M. Mooij, J.~Peters, D.~Janzing, J.~Zscheischler, B.~Sch{\"o}lkopf, \href{https://dl.acm.org/doi/10.5555/2946645.2946677}{Distinguishing Cause From Effect Using Observational Data: Methods and Benchmarks}, Journal of Machine Learning Research 17~(1) (2016) 1103--1204.

\bibitem{spearman2010proof}
C.~Spearman, \href{https://doi.org/10.1093/ije/dyq191}{The Proof and Measurement of Association Between Two Things}, International Journal of Epidemiology 39~(5) (2010) 1137--1150.

\bibitem{kendall1938new}
M.~G. Kendall, \href{https://doi.org/10.1093/biomet/30.1-2.81}{A New Measure of Rank Correlation}, Biometrika 30~(1/2) (1938) 81--93.

\bibitem{salvadori2007extremes}
G.~Salvadori, C.~De~Michele, N.~T. Kottegoda, R.~Rosso, \href{http://dx.doi.org/10.1007/1-4020-4415-1}{Extremes in Nature: An Approach Using Copulas}, Springer, 2007.

\bibitem{durante2016principles}
F.~Durante, C.~Sempi, \href{https://doi.org/10.1201/b18674}{Principles of Copula Theory}, CRC Press, 2016.

\bibitem{cherubini2004copula}
U.~Cherubini, E.~Luciano, W.~Vecchiato, \href{https://doi.org/10.1002/9781118673331}{Copula Methods in Finance}, John Wiley \& Sons, 2004.

\bibitem{salmon2009recipe}
F.~Salmon, \href{https://doi.org/10.1111/j.1740-9713.2012.00538.x}{Recipe for Disaster: The Formula That Killed Wall Street}, Wired Magazine 17~(3) (2009) 17--03.

\bibitem{mackenzie2014formula}
D.~MacKenzie, T.~Spears, \href{https://www.jstor.org/stable/43284238}{The Formula That Killed Wall Street: The Gaussian Copula and Modelling Practices in Investment Banking}, Social Studies of Science 44~(3) (2014) 393--417.

\bibitem{renard2007use}
B.~Renard, M.~Lang, \href{https://doi.org/10.1016/j.advwatres.2006.08.001}{Use of a Gaussian Copula for Multivariate Extreme Value Analysis: Some Case Studies in Hydrology}, Water Resources 30~(4) (2007) 897--912.

\bibitem{zhang2019copulas}
L.~Zhang, V.~P. Singh, \href{http://dx.doi.org/10.1017/9781108565103.019}{Copulas and Their Applications in Water Resources Engineering}, Cambridge University Press, 2019.

\bibitem{kumar2019copula}
P.~Kumar, \href{https://doi.org/10.1007/978-981-13-0872-7}{Copula Functions and Applications in Engineering}, in: Logistics, Supply Chain and Financial Predictive Analytics, Springer, 2019, pp. 195--209.

\bibitem{takeuchi2010constructing}
T.~T. Takeuchi, \href{http://dx.doi.org/10.1111/j.1365-2966.2010.16778.x}{Constructing a Bivariate Distribution Function With Given Marginals and Correlation: Application to the Galaxy Luminosity Function}, Monthly Notices of the Royal Astronomical Society 406~(3) (2010) 1830--1840.

\bibitem{zheng2021copula}
J.~Zheng, A.~D'Amour, A.~Franks, \href{https://www.afranks.com/publication/2021-zheng-copula/}{Copula-based Sensitivity Analysis for Multi-Treatment Causal Inference with Unobserved Confounding}, arXiv preprint arXiv:2102.09412 (2021).

\bibitem{zheng2022sensitivity}
J.~Zheng, J.~Wu, A.~D'Amour, A.~Franks, \href{https://doi.org/10.48550/arXiv.2208.06552}{Sensitivity to Unobserved Confounding in Studies with Factor-structured Outcomes}, arXiv preprint arXiv:2208.06552 (2022).

\bibitem{kruskal1958ordinal}
W.~H. Kruskal, \href{https://doi.org/10.2307/2281954}{Ordinal Measures of Association}, Journal of the American Statistical Association 53~(284) (1958) 814--861.

\bibitem{meyer2013bivariate}
C.~Meyer, \href{https://doi.org/10.1080/03610926.2011.611316}{The Bivariate Normal Copula}, Communications in Statistics-Theory and Methods 42~(13) (2013) 2402--2422.

\bibitem{Papamakarios2017MAF}
G.~Papamakarios, I.~Murray, T.~Pavlakou, \href{https://dl.acm.org/doi/10.5555/3294771.3294994}{Masked Autoregressive Flow for Density Estimation}, in: Neural Information Processing Systems (NeurIPS), 2017, pp. 2338--2347.

\bibitem{papamakarios2021NF_pmi}
G.~Papamakarios, E.~Nalisnick, D.~J. Rezende, S.~Mohamed, B.~Lakshminarayanan, \href{https://dl.acm.org/doi/abs/10.5555/3546258.3546315}{Normalizing Flows for Probabilistic Modeling and Inference}, Journal of Machine Learning Research 22~(57) (2021) 1--64.

\bibitem{kobyzev2020NF}
I.~Kobyzev, S.~Prince, M.~Brubaker, \href{https://doi.org/10.1109/TPAMI.2020.2992934}{Normalizing Flows: An Introduction and Review of Current Methods}, IEEE Transactions on Pattern Analysis and Machine Intelligence 43~(11) (2021) 3964--3979.

\bibitem{wehenkel2019UMNN}
A.~Wehenkel, G.~Louppe, \href{https://papers.nips.cc/paper/2019/hash/2a084e55c87b1ebcdaad1f62fdbbac8e-Abstract.html}{Unconstrained Monotonic Neural Networks}, in: Neural Information Processing Systems (NeurIPS), 2019, pp. 1545--1555.

\bibitem{Huang2018NAF}
C.~Huang, D.~Krueger, A.~Lacoste, A.~C. Courville, \href{http://proceedings.mlr.press/v80/huang18d.html}{Neural Autoregressive Flows}, in: International Conference on Machine Learning (ICML), 2018, pp. 2083--2092.

\bibitem{Wand1994}
M.~P. Wand, M.~C. Jones, Kernel Smoothing, CRC Press, 1994.

\bibitem{hoover2006causality}
K.~D. Hoover, \href{https://dx.doi.org/10.2139/ssrn.930739}{Causality in Economics and Econometrics}, SSRN eLibrary (2006).

\bibitem{DAOUD2024102973}
A.~Daoud, F.~D. Johansson, {The Impact of Austerity on Children: Uncovering Effect Heterogeneity by Political, Economic, and Family Factors in Low- and Middle-income Countries}, Social Science Research 118 (2024) 102973.

\bibitem{Daoud2017IMFchildhealth}
A.~Daoud, E.~Nosrati, B.~Reinsberg, A.~E. Kentikelenis, T.~H. Stubbs, L.~P. King, \href{https://doi.org/10.1073/pnas.1617353114}{Impact of International Monetary Fund programs on child health}, PNAS 114~(25) (2017) 6492--6497.

\bibitem{hallerod2013badgovernance_poorchildren}
B.~Haller{\"o}d, B.~Rothstein, A.~Daoud, S.~Nandy, \href{https://doi.org/10.1016/j.worlddev.2013.03.007}{Bad Governance and Poor Children: A Comparative Analysis of Government Efficiency and Severe Child Deprivation in 68 Low-and Middle-income Countries}, World Development 48 (2013) 19--31.

\bibitem{balgi2022counterfactual}
S.~Balgi, J.~M. Pe{\~n}a, A.~Daoud, {Counterfactually-Equivalent Structural Causal Modelling Using Causal Graphical Normalizing Flows}, in: International Conference on Probabilistic Graphical Models (PGM), 2024, pp. 164--181.

\bibitem{blau1967american}
P.~M. Blau, Duncan, The American Occupational Structure, John Wiley and Sons, 1967.

\bibitem{daoud2018structuraladjustments}
A.~Daoud, B.~Reinsberg, {{Structural Adjustment, State Capacity and Child Health: Evidence From IMF Programmes}}, Epidemiology 48~(2) (2018) 445--454.

\bibitem{DAOUD2019foodagri}
A.~Daoud, B.~Reinsberg, A.~E. Kentikelenis, T.~H. Stubbs, L.~P. King, \href{https://doi.org/10.1016/j.foodpol.2019.01.005}{The International Monetary Fund’s Interventions in Food and Agriculture: An Analysis of Loans and Conditions}, Food Policy 83 (2019) 204--218.

\bibitem{Balgi2025}
S.~Balgi, A.~Daoud, J.~M. Pe\~na, G.~Wodtke, J.~Zhou, {Deep Learning With DAGs}, Sociological Methods and Research (2025).

\end{thebibliography}

\end{document}